\documentclass[preprint,12pt]{article}
\pdfoutput=1
\usepackage{amsmath,amssymb,array,calc,rotating,epsfig,psfrag, amscd, datetime}
\usepackage{graphicx}
\numberwithin{equation}{section}
\usepackage[all]{xy}

\usepackage{color}
\usepackage[
      colorlinks=true,
      urlcolor=blue,    
      filecolor=blue,     
      citecolor=red,
      pdfstartview=FitV,
       bookmarksopen=true    
      ]{hyperref}
\usepackage[left=2cm,top=1cm,right=3cm,nohead]{geometry}

\newcommand{\nc}{\newcommand}

\definecolor{cardinal}{rgb}{0.6,0,0}
\definecolor{darkgreen}{rgb}{0,0.5,0}
\definecolor{golden}{rgb}{0.92, 0.7, 0}
\definecolor{midnight}{rgb}{0, 0, 0.5}
\definecolor{darkblue}{rgb}{0.2, 0, 0.8}

\nc{\ra}{\rightarrow} 
\nc{\lra}{\leftrightarrow} 
\nc{\Ra}{\Rightarrow} 
\nc{\LRa}{\Leftightarrow} 
\nc{\blp}{{\big (}}
\nc{\brp}{{\big )}}
\nc{\Blp}{{\Big (}}
\nc{\Brp}{{\Big )}}
\nc{\bglp}{{\bigg (}}
\nc{\bgrp}{{\bigg )}}
\nc{\Bglp}{{\Bigg (}}
\nc{\Bgrp}{{\Bigg )}}
\nc{\slb}{{\rm [}}
\nc{\srb}{{\rm ]}}
\nc{\bslb}{{\rm \big [}}
\nc{\bsrb}{{\rm \big ]}}
\nc{\Bslb}{{\rm \Big [}}
\nc{\Bsrb}{{\rm \Big ]}}

\def\al{\alpha}

\def\eps{\epsilon}
\nc{\veps}{\varepsilon}
\def\gam{\gamma}

\def\lam{\lambda}
\def\om{\omega}

\nc{\vphi}{\varphi}
\def\tha{\theta}

\def\sig{\sigma}

\def\Gam{\Gamma}

\def\Lam{\Lambda}
\def\Om{\Omega}
\def\Sig{\Sigma}

\def\coeff#1#2{\relax{\textstyle {#1 \over #2}}\displaystyle}

\nc{\myvspace}{\rule[-1em]{0pt}{2.5em}}
\nc{\bea}{\begin{eqnarray}}
\nc{\eea}{\end{eqnarray}}
\nc{\be}{\begin{equation}}
\nc{\ee}{\end{equation}}
\nc{\barr}{\begin{array}}
\nc{\earr}{\end{array}}

\nc{\cA}{{\cal A}}
\nc{\cB}{ \cal B}

\def\cD{{\cal D}}

\nc{\cF}{{\cal F}}
\nc{\cG}{{\cal G}}

\def\cI{{\cal I}}

\def\cK{{\cal K}}
\nc{\cL}{{\cal L}}
\nc{\cM}{{\cal M}}
\def\N{{\cal N}}
\def\cN{{\cal N}}

\nc{\cQ}{{\cal Q}}
\nc{\cR}{{\cal R}}
\def\cS{{\cal S}}

\def\cU{{\cal U}}
\def\cV{{\cal V}}
\def\cV{{\cal V}}

\def\cZ{{\cal Z}}
\nc{\cQd}{\cQ^{\dagger}}
\nc{\cRd}{\cR^{\dagger}}
\nc{\BB}{{\mathbb B}}
\nc{\CC}{{\mathbb C}}
\nc{\DD}{{\mathbb D}}
\nc{\EE}{{\mathbb E}}
\nc{\FF}{{\mathbb F}}
\nc{\GG}{{\mathbb G}}
\nc{\HH}{{\mathbb H}}
\nc{\JJ}{{\mathbb J}}
\nc{\MM}{{\mathbb M}}
\nc{\RR}{{\mathbb R}}
\nc{\PP}{{\mathbb P}}
\nc{\QQ}{{\mathbb Q}}
\nc{\UU}{{\mathbb U}}
\nc{\ZZ}{{\mathbb Z}}
\nc{\calone}{{\mathbb 1}}

\nc{\half}{\coeff{1}{2}}
\nc{\quarter}{\coeff{1}{4}}
\nc{\del}{\partial}

\nc{\delbar}{\bar\partial}
\nc{\thalf}{\frac{t}{2}}
\nc{\Spin}{\operatorname{Spin}}
\nc{\SO}{\operatorname{SO}}

\nc{\Sp}{{\rm Sp}}
\nc{\com}[2]{{ \left[ #1, #2 \right] }}
\nc{\acom}[2]{{ \left\{ #1, #2 \right\} }}
\nc{\rr}{\rightarrow}
\nc{\p}{\partial}
\nc{\LT}{{\LL_\T}}
\nc{\Tr}{{\rm Tr}}
\nc{\tr}{{\rm tr}}
\nc{\Adag}{A^{\dagger}}
\nc{\AdagI}{A^{\dagger I}}
\nc{\AdagJ}{A^{\dagger J}}
\nc{\AdagK}{A^{\dagger K}}
\nc{\AdagL}{A^{\dagger L}}
\nc{\AdagM}{A^{\dagger M}}
\nc{\Bdag}{B^{\dagger}}
\nc{\BdagI}{B^{\dagger}_I}
\nc{\BdagJ}{B^{\dagger}_J}
\nc{\BdagK}{B^{\dagger}_K}
\nc{\BdagL}{B^{\dagger}_L}
\nc{\BdagM}{B^{\dagger}_M}
\nc{\Cdag}{C^{\dagger}}
\nc{\CdagI}{C^{\dagger I}}
\nc{\CdagJ}{C^{\dagger J}}
\nc{\CdagK}{C^{\dagger K}}
\nc{\Ddag}{D^{\dagger}}
\nc{\DdagI}{D^{\dagger I}}
\nc{\DdagJ}{D^{\dagger J}}
\nc{\DdagK}{D^{\dagger K}}
\nc{\ttha}{\tilde{\theta}}
\nc{\ttau}{\tilde{\tau}}
\nc{\tTha}{\tilde{\Theta}}
\nc{\tphi}{\tilde{\phi}}
\nc{\tsig}{\tilde{\sig}}
\nc{\tom}{\widetilde{\om}}
\nc{\tOm}{\widetilde{\Om}}
\nc{\tlam}{\widetilde{\lam}}
\nc{\tLam}{\tilde{\Lam}}
\nc{\tSig}{\widetilde{\Sig}}
\nc{\tPhi}{\tilde{\Phi}}
\nc{\tPhibar}{\ol{\tPhi}}
\nc{\tPi}{\widetilde{\Pi}}
\nc{\tpsi}{\widetilde{\psi}}
\nc{\tPsi}{\tilde{\Psi}}
\nc{\tgam}{\widetilde{\gam}}
\nc{\tGam}{\widetilde{\Gam}}
\nc{\tzeta}{\tilde{\zeta}}
\nc{\tZeta}{\tilde{\Zeta}}
\nc{\teta}{\widetilde{\eta}}
\nc{\teps}{\tilde{\eps}}
\nc{\tveps}{\tilde{\veps}}
\nc{\tEta}{\tilde{\Eta}}
\nc{\tchi}{\tilde{\chi}}
\nc{\tChi}{\tilde{\Chi}}
\nc{\txi}{\tilde{\xi}}
\nc{\tXi}{\widetilde{\Xi}}
\nc{\tnu}{\tilde{\nu}}
\nc{\tmu}{\tilde{\mu}}

\nc{\tb}{\tilde b}
\nc{\tc}{\tilde c}
\nc{\te}{\tilde e}
\nc{\tf}{\tilde f}
\nc{\tg}{\tilde g}
\nc{\ti}{\tilde i}
\nc{\tj}{\tilde j}
\nc{\tk}{\tilde k}
\nc{\tl}{\tilde l}
\nc{\tm}{\tilde m}
\nc{\tn}{\tilde n}
\nc{\tp}{\tilde{p}}
\nc{\tq}{\widetilde{q}}
\nc{\ts}{{\tilde s}}
\nc{\tu}{{\tilde u}}
\nc{\tv}{{\tilde v}}
\nc{\tw}{{\tilde w}}
\nc{\tx}{{\tilde x}}
\nc{\ty}{{\tilde y}}
\nc{\tz}{\tilde z}
\nc{\tA}{{\widetilde A}}
\nc{\tAbar}{{\ol \tA}}
\nc{\tB}{{\widetilde B}}
\nc{\tC}{{\widetilde C}}
\nc{\tD}{{\widetilde D}}
\nc{\tE}{{\widetilde E}}
\nc{\tF}{{\widetilde F}}
\nc{\tG}{{\widetilde G}}
\nc{\tH}{{\widetilde H}}
\nc{\tJ}{{\widetilde J}}
\nc{\tJbar}{{\ol {\tilde J}}}
\nc{\tK}{{\widetilde K}}
\nc{\tL}{{\widetilde L}}
\nc{\tcL}{{\widetilde \cL}}
\nc{\tM}{{\widetilde M}}
\nc{\tN}{{\widetilde N}}
\nc{\tcN}{{\widetilde \cN}}
\nc{\tP}{{\widetilde P}}
\nc{\tQ}{{\widetilde Q}}
\nc{\tR}{{\widetilde R}}
\nc{\tS}{\widetilde{S}}
\nc{\tT}{\widetilde{T}}
\nc{\tU}{\widetilde{U}}
\nc{\tV}{\widetilde{V}}
\nc{\tW}{\widetilde{W}}
\nc{\tcF}{\widetilde{{\cal F}}}
\nc{\tX}{\widetilde{X}}
\nc{\tY}{\widetilde{Y}}
\nc{\tcZ}{\tilde{\cZ}}
\nc{\tcZbar}{\ol{\tcZ}}

\nc{\ha}{\hat a}
\nc{\hb}{\hat b}
\nc{\hc}{\widehat c}
\nc{\hd}{\widehat d}
\nc{\he}{\widehat e}
\nc{\hf}{\widehat f}
\nc{\hg}{\widehat g}
\nc{\hh}{\widehat h}
\nc{\hm}{\widehat m}
\nc{\hn}{\widehat n}
\nc{\hp}{\widehat p}
\nc{\hr}{\widehat r}
\nc{\hs}{\widehat s}
\nc{\hv}{\widehat v}
\nc{\hw}{\widehat w}
\nc{\hx}{\widehat x}
\nc{\hy}{\widehat y}
\nc{\hz}{\widehat z}
\nc{\zhat}{\hat z}
\nc{\hA}{\widehat{A}}
\nc{\hB}{\widehat{B}}
\nc{\hC}{\widehat{C}}
\nc{\hD}{\widehat{D}}
\nc{\hE}{\widehat{E}}
\nc{\hF}{\widehat{F}}
\nc{\hcF}{\widehat{\cF}}
\nc{\hG}{\widehat{G}}
\nc{\hH}{\widehat{H}}
\nc{\hJ}{\widehat{J}}
\nc{\hK}{\widehat{K}}
\nc{\hL}{\widehat{L}}
\nc{\hcL}{\widehat{\cL}}
\nc{\hM}{\widehat M}
\nc{\hcM}{\widehat{\cM}}
\nc{\hN}{\widehat{N}}
\nc{\hO}{\widehat{O}}
\nc{\hP}{\widehat{P}}
\nc{\hQ}{\widehat{Q}}
\nc{\hcR}{\widehat{\cR}}
\nc{\hR}{\widehat{R}}
\nc{\hS}{\widehat{S}}
\nc{\hcS}{\widehat{\cS}}
\nc{\hT}{\widehat{T}}
\nc{\hU}{\widehat{U}}
\nc{\hV}{\widehat V}
\nc{\hcV}{\widehat \cV}
\nc{\hX}{\widehat X}
\nc{\hY}{\widehat Y}
\nc{\hZ}{\widehat Z}
\nc{\hcZ}{\widehat{\cal Z}}

\nc{\heta}{\widehat{\eta}}
\nc{\hal}{\widehat \alpha}
\nc{\hphi}{\widehat{\phi}}
\nc{\hkap}{\hat{\kappa}}
\nc{\hchi}{\widehat{\chi}}
\nc{\hpsi}{\widehat{\psi}}
\nc{\hsig}{\widehat{\sig}}
\nc{\hgam}{\widehat{\gam}}
\nc{\hPhi}{\hat{\Phi}}
\nc{\hPsi}{\hat{\Psi}}
\nc{\hGam}{\hat{\Gam}}
\nc{\omhat}{\widehat{\om}}
\nc{\htha}{\hat{\tha}}
\nc{\hrho}{\widehat{\rho}}

\nc{\w}{\wedge}

\nc{\vb}{\vec b}
\nc{\vc}{\vec c}
\nc{\vd}{\vec d}
\nc{\ve}{\vec e}
\nc{\vf}{\vec f}
\nc{\vg}{\vec g}
\nc{\vh}{\vec h}
\nc{\vp}{\vec p}
\nc{\vq}{\vec q}
\nc{\vr}{\vec r}
\nc{\vs}{\vec s}
\nc{\vv}{\vec v}
\nc{\vw}{\vec w}
\nc{\vx}{\vec x}
\nc{\vy}{\vec y}
\nc{\vz}{\vec z}

\nc{\vB}{\vec B}
\nc{\vC}{\vec C}
\nc{\vD}{\vec D}
\nc{\vE}{\vec E}
\nc{\vF}{\vec F}
\nc{\vG}{\vec G}
\nc{\vH}{\vec H}
\nc{\vP}{\vec P}
\nc{\vQ}{\vec Q}
\nc{\vR}{\vec R}
\nc{\vS}{\vec S}
\nc{\vV}{\vec V}
\nc{\vW}{\vec W}
\nc{\vX}{\vec X}
\nc{\vY}{\vec Y}
\nc{\vZ}{\vec Z}

\nc{\ol}{\overline}
\nc{\abar}{\ol{a}}
\nc{\bbar}{\ol{b}}
\nc{\cbar}{\ol{c}}
\nc{\dbar}{\ol{d}}
\nc{\ebar}{\ol{e}}
\nc{\fbar}{\ol{f}}
\nc{\ibar}{\ol{\imath}}
\nc{\jbar}{\ol{\jmath}}
\nc{\kbar}{\ol{k}}
\nc{\lbar}{\ol{l}}
\nc{\mbar}{\ol{m}}
\nc{\nbar}{\ol{n}}
\nc{\pbar}{\ol{p}}
\nc{\qbar}{\ol{q}}
\nc{\rbar}{\ol{r}}
\nc{\sbar}{\ol{s}}
\nc{\ubar}{\ol{u}}
\nc{\vbar}{\ol{v}}
\nc{\wbar}{\ol{w}}
\nc{\xbar}{\ol{x}}
\nc{\ybar}{\ol{y}}
\nc{\zbar}{\ol{z}}

\nc{\Abar}{\ol{A}}
\nc{\Bbar}{\ol{B}}
\nc{\Cbar}{\ol{C}}
\nc{\Dbar}{\ol{D}}
\nc{\Ebar}{\ol{E}}
\nc{\Fbar}{\ol{F}}
\nc{\Jbar}{\ol{J}}
\nc{\Kbar}{\ol{K}}
\nc{\Lbar}{\ol{L}}
\nc{\cLbar}{\ol{\cL}}
\nc{\Mbar}{\ol{M}}
\nc{\Nbar}{\ol{N}}
\nc{\Pbar}{\ol{P}}
\nc{\Qbar}{\ol{Q}}
\nc{\Rbar}{\ol{R}}
\nc{\Sbar}{\ol{S}}
\nc{\Tbar}{\ol{T}}
\nc{\Ubar}{\ol{U}}
\nc{\Vbar}{\ol{V}}
\nc{\cVbar}{\ol{\cV}}
\nc{\Wbar}{\ol{W}}
\nc{\Xbar}{{\overline X}}
\nc{\Ybar}{{\overline Y}}
\nc{\Zbar}{{\overline Z}}
\nc{\cZbar}{{\overline \cZ}}

\nc{\epsbar}{\ol{\epsilon}}
\nc{\lambar}{\ol{\lambda}}
\nc{\kapbar}{\ol{\kappa}}
\nc{\zetabar}{\ol{\zeta}}
\nc{\Zetabar}{\ol{\Zeta}}
\nc{\taubar}{\ol{\tau}}
\nc{\Taubar}{\ol{\Tau}}
\nc{\psibar}{\ol{\psi}}
\nc{\Psibar}{\ol{\Psi}}
\nc{\tpsibar}{\ol{\tpsi}}
\nc{\tPsibar}{\ol{\tPsi}}
\nc{\phibar}{\ol{\phi}}
\nc{\Phibar}{\ol{\Phi}}
\nc{\chibar}{\ol{\chi}}
\nc{\mubar}{\ol{\mu}}
\nc{\nubar}{\ol{\nu}}
\nc{\rhobar}{\ol{\rho}}
\nc{\ombar}{\ol{\om}}
\nc{\Ombar}{\ol{\Om}}
\nc{\Deltabar}{\ol{\Delta}}
\nc{\Thetabar}{\ol{\Theta}}
\nc{\xibar}{\ol{\xi}}
\nc{\Xibar}{\ol{\Xi}}

\nc{\Dthbar}{\ol{\rm D3}}

\nc{\gdot}{\dot{g}}
\nc{\pdot}{\dot{p}}
\nc{\qdot}{\dot{q}}
\nc{\rdot}{\dot{r}}
\nc{\sdot}{\dot{s}}
\nc{\tdot}{\dot{t}}
\nc{\udot}{\dot{u}}
\nc{\vdot}{\dot{v}}
\nc{\wdot}{\dot{w}}
\nc{\xdot}{\dot{x}}
\nc{\xddot}{\ddot{x}}
\nc{\ydot}{\dot{y}}
\nc{\zdot}{\dot{z}}
\nc{\yddot}{\ddot{y}}

\nc{\Udot}{\dot{U}}
\nc{\Vdot}{\dot{V}}
\nc{\Wdot}{\dot{W}}

\nc{\taudot}{\dot{\tau}}
\nc{\phidot}{\dot{\phi}}
\nc{\psidot}{\dot{\psi}}
\nc{\sinp}{s_{\phi}}
\nc{\cosp}{c_{\phi}}
\nc{\tanp}{t_{\phi}}
\nc{\spone}{s_{\phi_1}}
\nc{\cpone}{c_{\phi_1}}
\nc{\tpone}{t_{\phi_1}}
\nc{\sptwo}{s_{\phi_2}}
\nc{\cptwo}{c_{\phi_2}}
\nc{\tptwo}{t_{\phi_2}}
\nc{\spth}{s_{\phi_3}}
\nc{\cpth}{c_{\phi_3}}
\nc{\tpth}{t_{\phi_3}}
\nc{\calp}{c_{\al}}
\nc{\salp}{s_{\al}}

\nc{\csch}{{\rm csch}}
\nc{\sech}{{\rm sech}}

\nc{\cothzlami}{\coth(z-\lam_i)}
\nc{\coshzlami}{\cosh(z-\lam_i)}
\nc{\sinhzlami}{\sinh(z-\lam_i)}

\nc{\cothzlamj}{\coth(z-\lam_j)}
\nc{\coshzlamj}{\cosh(z-\lam_j)}
\nc{\sinhzlamj}{\sinh(z-\lam_j)}

\nc{\cothlamij}{\coth(\lam_i-\lam_j)}
\nc{\coshlamij}{\cosh(\lam_i-\lam_j)}
\nc{\sinhlamij}{\sinh(\lam_i-\lam_j)}

\nc{\bah}{{\mathbf {\hat{A}}}}
\nc{\bX}{{\mathbf X}}
\nc{\ba}{{\bf a}}
\nc{\bb}{{\bf b}}
\nc{\bc}{{\bf c}}
\nc{\bd}{{\bf d}}
\nc{\bg}{{\bf g}}
\nc{\bk}{{\bf k}}
\nc{\bl}{{\bf l}}
\nc{\bm}{{\bf m}}
\nc{\bn}{{\bf n}}
\nc{\bo}{{\bf o}}
\nc{\bp}{{\bf p}}
\nc{\bq}{{\bf q}}
\nc{\br}{{\bf r}}
\nc{\bs}{{\bf s}}
\nc{\bt}{{\bf t}}
\nc{\bu}{{\bf u}}
\nc{\bv}{{\bf v}}
\nc{\bw}{{\bf w}}
\nc{\bx}{{\bf x}}
\nc{\by}{{\bf y}}
\nc{\bz}{{\bf z}}
\nc{\bom}{{\bf \om}}
\nc{\bombar}{{\mathbf \ombar}}
\nc{\bPhi}{{\bf \Phi}}

\nc{\rma}{{\rm a}}
\nc{\rmb}{{\rm b}}
\nc{\rmc}{{\rm c}}
\nc{\rmd}{{\rm d}}
\nc{\rmg}{{\rm g}}
\nc{\rk}{{\rm k}}
\nc{\rml}{{\rm l}}
\nc{\rmm}{{\rm m}}
\nc{\rmn}{{\rm n}}
\nc{\rmo}{{\rm o}}
\nc{\rmp}{{\rm p}}
\nc{\rmq}{{\rm q}}
\nc{\rmr}{{\rm r}}
\nc{\rms}{{\rm s}}
\nc{\rmt}{{\rm t}}
\nc{\rmu}{{\rm u}}
\nc{\rmv}{{\rm v}}
\nc{\rmw}{{\rm w}}
\nc{\rmx}{{\rm x}}
\nc{\rmy}{{\rm y}}
\nc{\rmz}{{\rm z}}

\nc{\dal}{\dot{\al}}
\nc{\thadot}{\dot{\tha}}
\nc{\thab}{\bar{\theta}}
\nc{\thal}{\theta^{\al}}
\nc{\thdal}{\bar{\theta}^{\dal}}

\nc{\thsigthm}{\tha \sigma^m \thab}
\nc{\thsigthn}{\tha \sigma^n \thab}

\nc{\Dal}{D_{\al}}
\nc{\Ddal}{\bar{D}_{\dal}}
\nc{\CDal}{{\cal D}_{\al}}
\nc{\CDdal}{\bar{\cal D}_{\dal}}

\nc{\eq}[1]{(\ref{#1})}
\nc{\non}{\nonumber}
\nc{\Fzero}{F_{(0)}}
\nc{\Ftwo}{F_{(2)}}
\nc{\Ffour}{F_{(4)}}
\nc{\Fone}{F_{(1)}}
\nc{\Fthree}{F_{(3)}}
\nc{\Ffive}{F_{(5)}}
\nc{\Fn}{F_{(n)}}
\nc{\Fp}{F_{(p)}}

\nc{\tFzero}{\tF_{(0)}}
\nc{\tFtwo}{\tF_{(2)}}
\nc{\tFfour}{\tF_{(4)}}
\nc{\tFone}{\tF_{(1)}}
\nc{\tFthree}{\tF_{(3)}}
\nc{\tFfive}{\tF_{(5)}}
\nc{\tFn}{\tF_{(n)}}
\nc{\tFp}{\tF_{(p)}}

\nc{\Czero}{C_{(0)}}
\nc{\Ctwo}{C_{(2)}}
\nc{\Cfour}{C_{(4)}}
\nc{\Cone}{C_{(1)}}
\nc{\Cthree}{C_{(3)}}
\nc{\Cfive}{C_{(5)}}
\nc{\Cn}{C_{(n)}}

\nc{\equ}{{\rm eq}}
\def\Im{{\rm Im \hspace{0.5mm} }}

\def\Re{{\rm Re \hspace{0.5mm}}}

\nc{\vol}{{\rm vol}}
\nc{\Ainf}{A_{\infty}}
\nc{\End}{{\rm End}}
\nc{\Ext}{{\rm Ext}}
\nc{\IIB}{{\rm IIB}}
\nc{\Ad}{{\rm Ad}}
\nc{\IIA}{{\rm IIA}}
\nc{\AdS}{{\rm AdS}}
\nc{\CFT}{{\rm CFT}}
\nc{\diag}{{\rm diag}}
\nc{\Log}{{\rm Log}}
\nc{\Dslash}{\ensuremath \raisebox{0.025cm}{\slash}\hspace{-0.32cm} D}
\nc{\cDslash}{\ensuremath \raisebox{0.025cm}{\slash}\hspace{-0.32cm} \cD}
\nc{\omslash}{\om\!\!\!/}
\nc{\no}{\!:\!\!}
\nc{\ointdz}{\oint\frac{dz}{2\pi i}}
\nc{\ointdzone}{\oint\frac{dz_1}{2\pi i}}
\nc{\ointdztwo}{\oint\frac{dz_2}{2\pi i}}
\nc{\ointdzb}{\oint\frac{d\zbar}{2\pi i}}
\nc{\ointdzbone}{\oint\frac{d\zbar_1}{2\pi i}}
\nc{\ointdzbtwo}{\oint\frac{d\zbar_2}{2\pi i}}
\nc{\dz}{\frac{dz}{2\pi i}}
\nc{\dzb}{\frac{d\zbar}{2\pi i}}
\nc{\bpm}{\begin{pmatrix}}
\nc{\epm}{\end{pmatrix}}
 \nc{\bitem}{\begin{itemize}}
 \nc{\eitem}{\end{itemize}}
 \nc{\exercise}{\vskip 2mm \noindent {\bf Exercise:}}
 \nc{\definition}{\vskip 2mm \noindent {\bf Definition:}}

\begin{document}

\vspace{0.5cm}
\begin{center}
\baselineskip=13pt {\LARGE \bf{BPS Black Holes in  $AdS_4$  \\
from M-theory\\}}
 \vskip1.5cm 
Nick Halmagyi$^*$, Michela Petrini$^*$, Alberto Zaffaroni$^{\dagger}$ \\ 
\vskip0.5cm
$^{*}$ Laboratoire de Physique Th\'eorique et Hautes Energies,\\
Universit\'e Pierre et Marie Curie, CNRS UMR 7589, \\
F-75252 Paris Cedex 05, France\\
\vskip0.5cm
$\dagger$ Dipartimento di Fisica, Universit\`a di Milano--Bicocca, \\
I-20126 Milano, Italy \\
and \\
INFN, sezione di Milano--Bicocca, \\
I-20126 Milano, Italy \\
\vskip0.5cm

halmagyi@lpthe.jussieu.fr \\ 
petrini@lpthe.jussieu.fr \\ 
alberto.zaffaroni@mib.infn.it
\end{center}

\begin{abstract}
We study supersymmetric black holes in $AdS_4$ in the framework of four dimensional gauged $\N=2$ supergravity coupled to hypermultiplets. We derive the flow equations for a general electrically gauged theory where the gauge group is Abelian and, restricting them to the fixed points, we derive the gauged supergravity analogue of the attractor equations for theories coupled to hypermultiplets. The particular models we analyze are consistent truncations of M-theory on certain Sasaki-Einstein seven-manifolds. We study the space of horizon solutions of the form $AdS_2\times \Sig_g$ with both electric and magnetic charges and find a four-dimensional solution space when the theory arises from a reduction on $Q^{111}$. For other $SE_7$ reductions, the solutions space is a subspace of this. We construct explicit examples of spherically symmetric black holes numerically.

\end{abstract}

\newpage

\section{Introduction}

Supersymmetric, asymptotically $AdS_4$ black holes\footnote{To be precise, the black holes we are discussing  will asymptotically approach $AdS_4$ in the UV but will differ by non-normalizable terms corresponding to some magnetic charge. We will nevertheless refer to them as asymptotically $AdS_4$ black holes.}   with regular spherical horizons have recently been discovered in $\N=2$ gauged supergravities with vector 
multiplets \cite{Cacciatori:2009iz}. These solutions have been 
 further studied in  \cite{DallAgata2011,Hristov:2010ri}.
 The analytic solution for the entire black hole was constructed and shown to be one quarter-BPS. 
For particular choices of prepotential and for particular values of the gauge couplings, these black holes can be embedded into M-theory and are asymptotic to $AdS_4\times S^7$. 

The goal of  this work is to study supersymmetric, asymptotically  $AdS_4$ black holes in more general gauged supergravities, with both vector and hypermultiplets. 
The specific theories we focus on are consistent truncations of string or M-theory. Supersymmetric black holes in these theories involve running hypermultiplet scalars and 
are substantially different from the examples in \cite{Cacciatori:2009iz}. The presence of hypers prevents us from finding analytic solutions of the BPS conditions, 
nevertheless we study analytically the space of supersymmetric horizon solutions $AdS_2\times \Sigma_g$  and  show that there is a large variety of them. We will then 
find explicit spherically symmetric black hole solutions interpolating between $AdS_4$ and $AdS_2\times S^2$ by numerical methods. The black holes we construct 
have both electric and magnetic charges.

 Our demand that the supergravity theory is a consistent truncation of M-theory and that the asymptotic $AdS_4$ preserves $\N=2$ supersymmetry  limits our search quite severely.  
Some of the gauged supergravity theories studied in \cite{Cacciatori:2009iz} correspond to the $\N=2$ truncations \cite{Cvetic1999b, Duff:1999gh} of the de-Wit/Nicolai 
$\N=8$ theory \cite{deWit:1981eq} where only massless vector multiplets are kept.  In this paper we will focus on more general theories obtained as consistent truncations of 
 M-theory on seven-dimensional Sasaki-Einstein manifolds.  
 A consistent truncation of  eleven-dimensional supergravity on a Sasaki-Einstein manifold to a universal sector was obtained in \cite{Gauntlett:2007ma, Gauntlett:2009zw}. 
 More  recently the  general reduction of eleven-dimensional supergravity  to four dimensions on left-invariant coset manifolds with  $SU(3)$-structure has been  performed in 
 \cite{Cassani:2012pj}\footnote{Other M-theory reductions have been studied in \cite{Donos:2010ax, Cassani:2011fu} and similar reductions have been performed in type IIA/IIB, 
 see for example \cite{Kashani-Poor:2006si, KashaniPoor:2007tr,Gauntlett:2010vu,Skenderis:2010vz, Cassani:2010uw, Liu:2010pq,Bena:2010pr, Cassani:2010na}}.  
 Exploiting the coset structure of the internal manifold it is possible to truncate the theory in such a way to also keep massive Kaluza-Klein multiplets.
These reductions can, by their very construction, be lifted directly to the higher dimensional theory and are guaranteed to solve the higher dimensional equations of motion. 

The black holes we construct represent the gravitational backreaction of bound states of M2 and M5-branes wrapped on curved manifolds in much the same manner as was 
detailed by Maldacena and Nunez \cite{Maldacena:2000mw} for D3-branes in $AdS_5 \times S^5$ and M5-branes in $AdS_7 \times S^4$. To preserve supersymmetry, a certain combination of the gauge connections in the bulk is set equal to the spin connection, having the  effect of twisting the worldvolume gauge theory in the manner of \cite{Witten:1988xj}. 
For D3-branes,  for particular charges, the bulk system will  flow to $AdS_3 \times \Sigma_g$  in the IR and the entire solution represents an asymptotically $AdS_5$ black string.
The general regular flow preserves just 2 real supercharges and thus in IIB string theory it is $\frac{1}{16}$-BPS.  Similarly,  for the asymptotically $AdS_7$, black M5-brane solutions,  depending on the charges, the IR geometry is $AdS_5\times \Sig_g$ and the dual $CFT_4$ may have $\N=2$ 
 or $\N=1$ supersymmetry. These $\N=2$ SCFT's and their generalizations have been of much recent interest \cite{Gaiotto2012h, Gaiotto2009} and the $\N=1$ case has also 
 been studied \cite{Benini:2009mz, Bah:2012dg}.

 By embedding the $AdS_4$   black holes in M-theory  we can see them as M2-brane wrapping a Riemann surface.  For  particular charges, the bulk
system will  flow to $AdS_2 \times \Sigma_g$  in the IR and represents a black hole with regular horizon. The original examples  found in   \cite{Caldarelli1999}
can be reinterpreted in this way;  it has  four   equal magnetic charges and can be embedded in $AdS_4 \times S^7$.  The explicit analytic solution is known
and it involves constant scalars and a hyperbolic horizon. A generalization of \cite{Maldacena:2000mw} to M2-branes wrapping $\Sig_g$ was performed in \cite{Gauntlett2002} where certain very symmetric twists were considered. Fully regular solutions for M2 branes wrapping a two-sphere with running scalars were finally found   in  \cite{Cacciatori:2009iz} in the form of $AdS_4$ 
black holes.  It is note-worthy that of all these scenarios of branes wrapping Riemann surfaces, 
 the complete analytic solution for general charges  is known  only for M2-branes on $\Sig_g$ with magnetic charges  \cite{Cacciatori:2009iz}. 
 
One way to generalize these constructions of branes wrapped on $\Sig_g$ is to have more general transverse spaces. This is  the focus of this article. For M5-branes one can orbifold $S^4$ while for D3-branes one can replace $S^5$ by an arbitrary $SE_5$ manifold and indeed a suitable consistent truncation on $T^{11}$ has indeed been 
constructed \cite{Bena:2010pr, Cassani:2010na}. For M2-branes one can replace $S^7$ by a seven-dimensional Sasaki-Einstein manifold $SE_7$ and, as discussed 
above, the work of \cite{Cassani:2012pj} provides us with a rich set of consistent truncations to explore.  Interestingly, in our analysis we find that there are no solutions for pure M2-brane backgrounds, there must be additional electric
and magnetic charges corresponding to wrapped M2 and M5-branes on internal cycles.  
Asymptotically $AdS_4$ black holes with more general transverse 
space can be found in \cite{Donos:2008ug} and \cite{Donos2012d} where the solutions were studied directly in M-theory.  These include the M-theory lift of the solutions we give in 
Sections \ref{sec:Q111Simp} and \ref{numericalQ111}.

The BPS black holes we construct in this paper are asymptotically $AdS_4$ and as such they are states in particular (deformed) three-dimensional superconformal 
field theories on $S^2\times \mathbb{R}$. 
The solution in  \cite{Cacciatori:2009iz} can be considered as a state in the twisted ABJM theory \cite{Aharony:2008ug}. The solutions we have found in this paper can be 
seen as states in (twisted and deformed) three dimensional Chern-Simons matter theory dual to the M-theory compactifications of homogeneous Sasaki-Einstein 
manifolds\footnote{For a discussion of these compactifications from the point of view of holography  and recent results in identifying the dual field theories 
see\cite{Fabbri:1999hw,Jafferis:2008qz,Hanany:2008cd,Martelli:2008rt,Hanany:2008fj,Martelli:2009ga,Franco:2009sp,Benini:2009qs}.}.  One feature of these theories 
compared to ABJM is the presence of many baryonic symmetries that couple to the vector multiplets arising from non trivial two-cycles in the Sasaki-Einstein manifold.   
In terms of the worldvolume  theory, the black holes considered in this paper are then electrically charged states  of a Chern-Simons matter theory in a monopole background 
for $U(1)_R$ symmetry  and other global  symmetries, including the baryonic ones\footnote{For a recent discussion
from the point of view of holography see \cite{Hristov:2013spa}.}.

Gauged $\N=2$ supergravity with hypermultiplets is the generic low-energy theory arising from a Kaluza-Klein reduction of string/M-theory on a flux background. 
The hypermultiplet scalars interact with the vector-multiplet scalars through the scalar potential: around a generic $AdS_4$ vacuum the eigenmodes mix the hypers and vectors. 
In the models we study, we employ a particular simplification on the hypermultiplet scalar manifold \eq{P120}  and find solutions where only one real hypermultiplet scalar 
has a non-trivial profile. Given that the simplification is so severe it is quite a triumph that solutions exist within this ansatz. It would be interesting to understand if this 
represents a general feature of black holes in gauged supergravity. \\

The paper is organized as follows. 
In Section 2 we summarize the ansatz we use  and the resulting BPS equations for an arbitrary electrically gauged $\N=2$ supergravity theory. The restriction of the flow 
equations to the horizon produces gauged supergravity analogues of the attractor equations. 

In Section 3 we describe the explicit supergravity models we consider.
A key step is that we use a symplectic rotation to a frame where the gauging parameters are purely electric so that we can use  the  supersymmetry variations 
at our disposal.  

In Section 4 we study horizon geometries of the form $AdS_2\times \Sig_g$ where $g\neq 1$. We find a four parameter solution space for $Q^{111}$ and the solutions spaces for all the other models are truncations of this space.

In Section 5 we construct numerically black hole solutions for $Q^{111}$ and for $M^{111}$. The former solution is a gauged supergravity reproduction of the solution 
found in \cite{Donos2012d} and is distinguished in the space of all solutions by certain simplifications. For this solution, the phase of the four dimensional spinor is constant 
and in addition the massive vector field vanishes. The solution which we construct in $M^{111}$ turns out to be considerably more involved to compute numerically and has 
all fields of the theory running. In this sense we believe it to be representative of the full solution space in $Q^{111}$.

\section{The Black Hole Ansatz}

We want to study  static supersymmetric asymptotically $AdS_4$ black holes in four-dimensional $\mathcal{N}=2$ gauged supergravity.  The standard conventions and notations for $\mathcal{N}=2$ gauged supergravity  \cite{Andrianopoli:1996vr,Andrianopoli:1996cm}  are briefly reviewed in Appendix  \ref{gsugra}. 

Being supersymmetric, these black holes can be found by solving the supersymmetry variations \eqref{gravitinoeq} - \eqref{hyperinoeq} plus Maxwell equations \eqref{Maxwelleq}.  In this section we give  the ansatz for the metric and the gauge fields, and a simplified form of the SUSY variations we will study in the rest of this paper. The complete SUSY variations are derived and discussed in Appendix  \ref{sec:BPSEqs}.

\subsection{The Ansatz}\label{sec:bhansatz}

We will focus on asymptotically $AdS_4$ black holes with spherical ($AdS_2\times S^2$)  or hyperbolic ($AdS_2\times \HH^2$) horizons.
The modifications required to study  $AdS_2\times \Sig_g$  horizons, where $\Sigma_g$ is a Riemann surface of genus $g$, are discussed at the end of Section \ref{sec:BPSflow}.
The ansatz for the metric and gauge fields is
\bea\label{ansatz}
ds^2&=& e^{2U} dt^2- e^{-2U} dr^2- e^{2(V-U)} (d\tha^2+F(\tha)^2 d\vphi^2) \label{metAnsatz}\\
A^\Lam&=& \tq^\Lam(r) dt- p^\Lam(r) F'(\tha)  d\vphi \,, 
\eea
with
\be
F(\tha)=\left\{ \barr{ll} \sin \tha:& S^2\ (\kappa=1) \\ 
\sinh \tha:& \HH^2\ (\kappa=-1) \earr \right. 
\ee

The electric and magnetic charges are
\bea
p^\Lam &=& \frac{1}{4\pi } \int_{S^2} F^\Lam \label{elinv} \, , \\
 q_\Lam &\equiv& \frac{1}{4\pi} \int_{S^2} G_\Lam = -e^{2(V-U)} \cI_{\Lam\Sig} \tq'^\Sig +\cR_{\Lam\Sig} \kappa p^\Sig\, , 
  \label{maginv}
\eea
where $G_\Lam$ is the  symplectic-dual gauge field strength 
\be
G_{\Lam} \equiv \frac{\delta \cL}{ \delta F^\Lam}=R_{\Lam \Sig} F^\Lam -\cI_{\Lam \Sig} *F^\Sig \, . 
\ee
In addition,  we assume that all  scalars in the theory,  the fields $z^i$ from the $n_v$-vector multiplets and $q^u$ from the $n_h$-hypermultiplets, 
are  functions of the radial coordinate $r$, only.  Moreover, we will restrict our analysis to  abelian gaugings of the hypermultiplet moduli space and assume
that the gauging is purely electric.  As discussed in  \cite{deWit:2005ub}, for Abelian gauge groups one can always find a symplectic frame where this is true. 

\subsection{The BPS Flow Equations}\label{sec:BPSflow}

In Appendix  \ref{sec:BPSEqs}, we derive the general form that the SUSY conditions take with our ansatz for the metric and gauge fields and  the hypothesis discuss above for the 
gaugings. We will only consider  spherical and hyperbolic horizons.  

Throughout the text, when looking for explicit black hole solutions we make one simplifying assumption, namely that the Killing prepotentials $P^x_\Lam$ of the hypermultiplet scalar manifold $\cM_h$  satisfy\footnote{For the models studied in this paper, this also implies $\omhat_\mu^x=0$ in \eq{DepsA}}
\be
P^1_\Lam=P^2_\Lam=0\,. \label{P120}
\ee
The flow equations given in this section reduce to the equations in \cite{DallAgata2011,Hristov:2010ri} when the hypermultiplets are truncated away and thus $P^3_\Lam$ are constant. \\

The preserved supersymmetry is
\be
\eps_A= e^{U/2} e^{i\psi/2} \eps_{0A}
\ee
where  $\eps_{0A}$ is an $SU(2)$-doublet of constant spinors  which satisfy the following projections
\bea
\eps_{0A}&=&i \eps_{AB}\gam^{0}  \eps_0^{B}\,,  \\
\eps_{0A}&=& \mp \,(\sigma^3)_A^{\ B} \gam^{01} \epsilon_{0B}  \,.
\eea
As a result only $2$ of the  8  supersymmetries are preserved along any given flow.  Imposing these two projections, the remaining content of the supersymmetry equations reduces 
to a set of bosonic  BPS equations. Some are algebraic 
\bea
p^\Lam P_\Lam^3&=&\pm 1 \label{pP1} \, , \\
p^\Lam k_{\Lam}^u  &=& 0 \label{pk1}  \, ,  \\
\cL_r^\Lam P^3_\Lam&=& \pm  e^{2(U-V)}\Im \blp   e^{-i\psi}\cZ \brp \label{Alg1}  \, ,  \\
\tq^\Lam P^3_\Lam  &=& 2 e^U \cL_r^{\Lam}P^3_\Lam \label{qP}  \, ,  \\
\tq^\Lam k^u_\Lam  &=& 2 e^U \cL_r^{\Lam}k^u_\Lam  \label{qk}  \, , 
\eea
and some differential 
\bea 
(e^U)'&=& \pm \cL_i^\Lam P^3_{\Lambda} - e^{2(U-V)}\Re( e^{-i\psi}  \cZ)  \label{UEq}  \, , \\
 V'  &=&  \pm2 e^{-U} \cL_i^\Lam P^3_{\Lambda}   \label{VEq}  \, , \\
z'^i   &=&  e^{i\psi}e^{U-2V}g^{i\ibar}D_{\ibar} \cZ 
 \mp i e^{i\psi}e^{-U}  g^{i\jbar} {\bar f}_{\jbar}^{\Lambda} P_\Lam^3 \label{tauEq}  \, , \\
  q'^u&=&\mp 2e^{-U} h^{uv} \del_v \Blp  \cL_i^{\Lam} P^3_\Lam\Brp  \label{qEq}  \, , \\
\psi'&=&-A_r\mp e^{-2U}\tq^\Lam P_\Lam^3   \label{psiEq}  \, , \\
p'^\Lam&=& 0  \, , 
\eea
where  we have absorbed a phase in the definition of the symplectic  sections 
\be
\cL^\Lam=\cL_r^\Lam+i \cL_i^\Lam= e^{-i\psi} L^\Lam \, .
\ee
$\cZ$ denotes  the central charge 
\bea
\cZ&=&p^\Lam M_\Lam- q_\Lam L^\Lam \non \\
&=& L^\Sig  \cI_{\Lam \Sig} (e^{2(V-U)} \tq^\Lam + i\kappa p^\Lam)\,, \\
D_{\ibar} \cZ &=&\fbar^{\Sigma}_{\ibar} \cI_{\Sig \Lam} \blp e^{2(V-U)} \tq'^{\Lam} +i\kappa p^{\Lam} \brp \,.
\eea
Once $P^3_\Lam$ are fixed, the $\pm$-sign in the equations above  can be absorbed by a redefinition  $(p^\Lam,q_\Lam,e^U)\ra -(p^\Lam,q_\Lam,e^U)$.\\

Since the gravitino and hypermultiplets are charged, there are standard Dirac quantization conditions which must hold in the vacua of the theory
\bea
p^\Lam P^3_\Lam &\in& \ZZ\,, \\
p^\Lam k^u_\Lam &\in& \ZZ\,.
\eea
We see from \eq{pP1} and \eq{pk1} that the BPS conditions select a particular integer quantization. \\

Maxwell's equation becomes 
\be
q'_\Lam =2 \, e^{-2U} e^{2(V-U)}h_{uv} k^u_\Lam k^v_\Sig \tq^\Sig \label{Max1} \, .
\ee
Notice that  for the truncations of M-theory studied in this work, the non-trivial RHS will play a crucial role since massive vector fields do not carry conserved charges. \\
 
Using standard special geometry relations, one can show that the variation for the vector multiplet  scalars and the warp factor $U$, \eq{UEq} and \eq{tauEq}, are
equivalent to a pair of constraints for the  sections $\cL^\Lam$  
\bea
\del_r  \blp e^{U} \cL_r^\Delta \brp  & =& \half   \tq'^{\Delta}   \, ,    \label{delLr2}\\
\del_r \blp e^{-U} \cL_i^\Delta \brp& =&\frac{  \kappa  p^{\Delta}}{2e^{2V}} \pm \frac{ 1}{2 e^{2U }}   \cI^{\Delta \Sig} P_\Sig^3   \pm 2 e^{-3U} \tq^\Delta P^3_\Delta \cL_r^\Lam  \, . 
 \label{delLi2}
\eea
Importantly we can integrate \eq{delLr2} to get
\be
\tq^\Lam =2 e^U \cL_r^\Lam + c^\Lam \label{qLr}
\ee
for some constant $c^\Lam$. From \eq{qP} and \eq{qk} we see that this gauge invariance is constrained to satisfy
\be
c^\Lam P^3_\Lam=0\,,\ \ \ \ c^\Lam k^u_\Lam=0\,.
\ee 
We note that  due to the constraint on the sections
\be
\cI_{\Lam \Sig} \cL^\Lam \cLbar^\Sig=-\frac{1}{2}\,,
\ee
\eq{delLr2} and \eq{delLi2} give $(2n_v+1)$-equations. \\

One can show that the algebraic relation \eq{Alg1} is an integral of motion for the rest of the system. Specifically, differentiating \eq{Alg1} one finds a combination of the BPS 
 equations plus  Maxwell equations contracted with $\cL_i^\Lam$. One can solve \eq{Alg1} for $\psi$ and find that it is the phase of a modified ``central charge" $\hcZ$:
\bea
\hcZ&=& e^{i\psi}|\hcZ|\,,\ \ \ \  \hcZ=(e^{2(U-V)}\cZ\mp i L^\Lam P^3_\Lam)\,.
\eea

\vspace{0.2cm}

Our analysis also applies to black holes with    $AdS_2\times \Sig_g$  horizons,  where $\Sig_g$ is a Riemann surface of genus $g\ge 0$. The case
$g>1$ is trivially obtained by taking a quotient of $\HH^2$ by a discrete group, since all Riemann surfaces with $g > 1$  can be obtained
in this way.  Our system of BPS equations (\ref{pP1}) - (\ref{psiEq})   also applies   to the case of flat or toroidal horizons ($g=1$) 
\bea
ds^2&=& e^{2U} dt^2- e^{-2U} dr^2- e^{2(V-U)} (d x^2+ d y^2) \\
A^\Lam&=& \tq^\Lam(r) dt- p^\Lam(r) x d y \,, 
\eea
with
\bea
 q_\Lam &\equiv& -e^{2(V-U)} \cI_{\Lam\Sig} \tq'^\Sig - \cR_{\Lam\Sig}  p^\Sig\, , \\
\cZ&=&L^\Sig  \cI_{\Lam \Sig} (e^{2(V-U)} \tq^\Lam  - i  p^\Lam)\,, 
\eea
provided we substitute the constraint (\ref{pP1}) with
\be
p^\Lam P_\Lam^3= 0 \, .
\ee
We will not consider explicitly the case of flat horizons in this paper although they have attracted some recent interest \cite{Donos2012d}.

\subsection{$AdS_2\times S^2$ and $AdS_2\times \HH^2$ Fixed Point Equations}\label{sec:horizonEqs}

At the horizon the scalars $(z^i,q^u)$ are constant, while the functions in the metric and gauge fields take the form 
\be
e^U=\frac{r}{R_1}\,,\ \ \ \ \  e^V=\frac{r R_2}{R_1}\,, \ \ \ \ \ \ \tq^\Lam = r\tq_0^\Lam\\
\ee
with $q_0^\Lam$  constant. 
The BPS equations are of course much simpler, in particular they  are all algebraic and there are additional superconformal symmetries.  

There are the two Dirac quantization conditions
\bea
p^\Lam P_\Lam^3 &=&\pm 1\,, \label{pP} \\
p^\Lam k_{\Lam}^u &=&0 \, ,  \label{pk}
\eea
and  \eqref{psiEq}  \eqref{Max1} give two constraints on the electric component of the gauge field
\bea
\tq_0^\Lam P^x_\Lam&=& 0\, ,  \label{tqP} \\
\tq_0^\Lam k^u_\Lam&=& 0\,  \label{tqk} \, . 
\eea
The radii are given by \eqref{UEq} and \eqref{VEq}
\bea
 \frac{1}{R_1} &=&\pm 2 \cL_i^\Lam P^3_\Lam \label{R1hor} \, , \\
 \frac{R_2^2}{R_1}&=&- 2\Re( e^{-i\psi}  \cZ) \label{R2hor} \, .
\eea
In addition,  the algebraic constraint \eq{Alg1} becomes 
\be
\Im (e^{-i\psi} \cZ)=0
\ee
and the hyperino variation gives
\be
\cL_i^\Lam k^u_\Lam =0\,.\label{hyphor}
\ee

Finally, combining    \eqref{qLr}, \eqref{delLi2} and \eqref{maginv}, we can express   the 
charges  in terms of the scalar fields 
\bea
\kappa  p^{\Lam}&=& -\frac{2R_2^2}{R_1} \cL_i^\Lam  \mp  R_2^2  \cI^{\Lam \Sig} P_\Sig^3 \label{phor} \, , \\
q_\Lam&=& -\frac{2R_2^2}{R_1} \cM_{i\,\Lam} \mp R_2^2\, \cR_{\Lam \Sig}  \cI^{\Sig \Delta} P^3_\Delta  \, ,  \label{qhor}
\eea
with  $\cM_{i\,\Lam}=\Im (e^{-i\psi} M_\Lam)$.
These are the gauged supergravity analogue of the {\it attractor equations}. \\

It is of interest to solve explicitly for the spectrum of horizon geometries in any given gauged supergravity theory.
In particular this should involve inverting \eq{phor} and \eq{qhor} to express the scalar fields in terms of the charges. Even in the ungauged case, this is  in general not possible analytically and the equations here are considerably more complicated. Nonetheless one can determine the dimension of the solution space  and, for any particular  set of charges,
 one can numerically solve the horizon equations to determine the value of the various scalars. In this way one can check regularity of the solutions.

\section{Consistent Truncations of M-theory} \label{sec:truncations}

Having massaged the BPS equations into a neat set of bosonic equations we now turn to particular gauged supergravity theories in order to analyze the space of black hole solutions. We want to study models which have consistent lifts to M-theory and which have an $\cN=2$ $\ AdS_4$ vacuum somewhere in their field space, this limits our search quite severely. Two examples known to us are $\N=2$ truncations of the de-Wit/Nicolai $\N=8$ theory \cite{deWit:1981eq} and the truncation of M-theory on $SU(3)$-structure cosets \cite{Cassani:2012pj}. 
In this paper we will concentrate on some of the models constructed in \cite{Cassani:2012pj}.  The ones of interest for us  are listed in Table 1. \\

\begin{table}[bth!]\label{tb1}
\begin{center}
\begin{tabular}{|c|c|c|c|} 
\hline
$M_7$& $n_v:m^2=0$&$n_v:m^2\neq0$& $n_h$ \\
\hline
$Q^{111}$ & 2& 1& 1\\
\hline
$M^{111}$ &1 &1 & 1\\
\hline
$N^{11}$ &1 &2 & 2\\
\hline
$\frac{Sp(2)}{Sp(1)}$ &0 &2 &2 \\
\hline
$\frac{SU(4)}{SU(3)}$ &0 &1 &1 \\
\hline
\end{tabular}
\caption{The consistent truncations on $SU(3)$-structure cosets being considered in this work. $M_7$ is the 7-manifold, the second column is the number of massless vector multiplets at the $AdS_4$ vacuum, the third column is the number of massive vector multiplets and final column is the number of hypermultiplets.}
\end{center}
\end{table}

For each of these models there exists a consistent truncation to an $\N=2$ gauged supergravity with $n_v$ vector multiplets and $n_h$ hypermultiplets.
We summarize here some of the features of these models referring to 
 \cite{Cassani:2012pj}  for a more detailed discussion.

We denote the vector  multiplets scalars 
\be
z^i = b^i + i v^i     \qquad i = 1, \ldots , n_v
\ee
where the number of vector multiplets $n_v$ can vary from 0 to  3.  Notice that  all  models  contain some massive vector multiplets. For the  hypermultiplets, we use the notation
\be
( z^i, a, \phi, \xi^A)
\ee
where $a, \phi$ belong to the universal hypermultiplet.  This is motivated by the structure of the quaternionic moduli spaces in these models, which 
 can be seen as images of the c-map. The metric on quaternionic K\"ahler manifolds of this kind can 
be written in the form  \cite{Ferrara:1989ik}
\be
ds_{QK}^2=d\phi^2 +g_{i\jbar} dz^i d\zbar^{\jbar} +\quarter e^{4\phi}\blp da+\half \xi^T \CC d \xi \brp^2 -\quarter e^{2\phi}d\xi^T \CC \MM d\xi\,,
\ee
where $\{z^i,\zbar^{\jbar}|i=1,\ldots,n_h-1\}$ are special coordinates on the special K\"ahler manifold $\cM_c$ and $\{ \xi^A,\txi_A| A=1,\ldots,n_h\}$ form the symplectic vector $\xi^T=(\xi^A,\txi_A)$ and are coordinates on the axionic fibers. \\

All these models, and more generally of $\mathcal{N}=2$ actions obtained from compactifications,  have a cubic prepotential for the vector multiplet scalars
and both magnetic and electric  gaugings  of abelian isometries of the hypermultiplet scalar manifold.  In ungauged supergravity the  vector multiplet sector is invariant under
$Sp(2n_v+2,\RR)$.  The gauging typically breaks this invariance,  and  we can use such an action to find a symplectic frame where the gauging is purely electric\footnote{This is always possible when the gauging is abelian  \cite{deWit:2005ub}.}.   Since  $Sp(2n_v+2,\RR)$ acts non trivially on the prepotential $\mathcal{F}$, the rotated models we study
will have a different prepotential than the original ones in \cite{Cassani:2012pj} .

\subsection{The Gaugings}
\label{sec:gaugings}

In the models we consider,  the symmetries of the hypermultiplet moduli space that are gauged are non compact shifts of the axionic fibers $\xi_A$ and
$U(1)$ rotations of the special K\"ahler basis $z^i$.  The corresponding  Killing vectors  are the Heisenberg vector fields:
\bea
h^A&=& \del_{\txi_A} + \half \xi^A \del_a\,,\ \ \ \ h_A= \del_{\xi^A} - \half \txi_A \del_a\,, \ \ \ \ h=\del_a \, 
\eea
which satisfy $[h_A,h^B]=\delta_A^B h$, as well as 
\bea
f^A&=&\txi_A\del_{\xi^A}- \xi^A\del_{\txi_A}\,, \ \ \ \ ({\rm indices\ not\ summed}) \\ 
g&=&\zbar \del_{z}+ z \del_{\zbar}   \,.
\eea	

\vspace{0.1cm}

For some purposes it is convenient to work in homogeneous coordinates on $\cM_c$
\be
\xi = \begin{pmatrix} \xi^A \\  \xi_A \end{pmatrix}   \qquad \qquad  Z =  \begin{pmatrix} Z^A \\  Z_A  \end{pmatrix}
\ee
with  $z^i = Z^i/Z^0$ and to define 
\be
k_\UU=(\UU Z)^A \frac{\del}{\del Z^A}+(\UU \Zbar)^A \frac{\del}{\del \Zbar^A}+(\UU \xi)^A \frac{\del}{\del \xi^A}+(\UU \xi)_A \frac{\del}{\del \txi_A} \, , 
\ee
where $\UU$ is a $2n_h\times 2n_h$ matrix of gauging parameters. In special coordinates $k_\UU$ is a sum of the Killing vectors $f^A$ and $g$.

A general electric  Killing vector field of the quaternionic K\"ahler manifold is given by
\be
k_\Lam =k^u_\Lam \frac{\del }{\del q^u}=\delta _{0\Lam} k_{\UU} +Q_{\Lam A} h^A+Q_{\Lam}^{\ A} h_A -e_{\Lam} h \, ,
\ee
where $Q_{\Lam A}$ and $Q_\Lam^A$ are also matrices of gauge parameters, while the magnetic gaugings are parameterized by  \cite{Cassani:2012pj} 
\be
\tk^{\Lam}=-m^\Lam h\,.
\ee
For these models, the resulting Killing prepotentials can be worked out using the property
\be
 \label{Pkw}
P^x_\Lam =k^u_\Lam \om^x_u \qquad \qquad \tilde{P}^{x \, \Lam} = \tilde{k}^{u \, \Lam}  \om^x_u \, , 
\ee
where $\om^x_u$ is the spin connection on the quaternionic K\"ahler manifold \cite{Ferrara:1989ik} 
\bea
\om^1+i \om^2&=& \sqrt{2}e^{\phi + K_{c}/2} Z^T \CC d\xi\,, \\
\om^3 &=&\frac{e^{2\phi}}{2} \blp da + \half \xi^T \CC d\xi \brp - 2 e^{K_c} \Im \Blp Z^A\Im \cG_{AB} d\Zbar^B \Brp\,.
\eea
The Killing vector $k_\UU$ may contribute a constant shift to $P^3_0$, and this is indeed the case for the examples below.

As already mentioned, we will work in a rotated frame where all gaugings are electric. The form of the Killing vectors and prepotentials is the same, with the only difference
that now $\tk^{\Lam}=-m^\Lam h$ and $ \tilde{P}^{x \, \Lam}$ will add an extra contribution to the electric ones. 

\subsection{The Models}

The models which we will study are summarized in Table 1.  They all  contain an $AdS_4$ vacuum with $\mathcal{N}=2$ supersymmetry. 
The vacuum corresponds to the ansatz (\ref{ansatz}) with warp factors
\be
e^U=\frac{r}{R}\,,\ \ \ \ e^{V}=\frac{r^2}{R}\,,
\ee
and no electric and magnetic charges   \be p^\Lam= q_\Lam= 0 \, .\ee 

The $AdS_4$  radius and the non trivial scalar fields are 
\be
R=\frac{1}{2}\Blp \frac{e_0}{6}\Brp^{3/4}\,, \ \ \ \  v_i= \sqrt{\frac{e_0}{6}}\,, \ \ \ \ e^{-2\phi} = \frac{e_0}{6}\,.\label{AdS4Sol}
\ee

This is not an exact solution of the flow equations in Section \ref{sec:BPSflow} which require a non-zero magnetic charge to satisfy \eq{pP1}. The black holes of this paper will asymptotically approach $AdS_4$ in the UV but will differ by non-normalizable terms corresponding to the magnetic charge. The corresponding asymptotic behavior has been dubbed {\it magnetic} $AdS$ in \cite{Hristov:2011ye}.

\subsubsection{$Q^{111}$}

The scalar manifolds for the $Q^{111}$ truncation are
\be
\cM_v=\Blp \frac{SU(1,1)}{U(1)}\Brp^3\,,\ \ \ \cM_h= \cM_{2,1} =\frac{SU(2,1)}{SU(2)\times U(1)} \, . 
\ee

The metric on $\cM_{2,1}$ is
\be
ds^2_{2,1}=d\phi^2 +\quarter e^{4\phi} \bslb da+\half(\xi^0 d\txi_0-\txi_0 d\xi^0) \bsrb^2 + \quarter e^{2\phi}\blp (d\xi^0)^2 + d\txi_0^2\brp \, , 
\ee
and the special K\"ahler base $\cM_c$ is trivial. Nonetheless we can formally use the prepotential and special coordinates on $\cM_c$
\be
\cG=\frac{(Z^0)^2}{2i}\,,\ \ \ \ Z^0=1
\ee
to construct the spin connection and Killing prepotentials. \\

The natural duality frame which arises upon reduction has a cubic prepotential\footnote{We slightly abuse notation by often refering to the components of $z^i$ as $(v_i,b_i)$. This is not meant to imply that the metric has been used to lower the index.}
\be
F=-\frac{X^1 X^2 X^3}{X^0} \label{FQ111} \, , \\
\ee
with sections $X^\Lam =  (1,z^)$ and  both electric and magnetic  gaugings 
\be
\UU = \bpm  0 & 4 \\-4 & 0\epm\,,\ \ \ \ e_0\neq0\,,\ m^1=m^2=m^3=-2\,. \label{gaugeQ111}
\ee
Using an element $\cS_0\in Sp(8,\ZZ)$ we rotate to a frame where the gaugings are purely electric. Explicitly we have
\be
\cS_0=\bpm A & B \\ C& D \epm\,,\ \ \ A=D= \diag(1,0,0,0)  \,,\ \ \  B=-C= \diag(0,-1,-1,-1)\label{S0rotation}
\ee
and the new gaugings are
\be
\UU = \bpm  0 & 4 \\-4 & 0\epm\,,\ \ \ \ e_0\neq0\,, \ e_1=e_2=e_3=-2\,. \label{gaugeQ11elec}
\ee
The Freund-Rubin parameter  $e_0>0$ is unfixed. In this duality frame the special geometry data are
\bea
F&=&2\sqrt{X^0X^1X^2X^3}\,, \\
X^\Lam&=& (1,z^2 z^3,z^1z^3, z^1 z^2)\,, \\
F_\Lam&=& (z^1z^2 z^3,z^1,z^2,z^3)\,.
\eea

\subsubsection{$M^{111}$ }

The consistent truncation on $M^{111}$ has
\be
\cM_v=\Blp \frac{SU(1,1)}{U(1)}\Brp^2\,,\ \ \ \cM_h= \cM_{2,1}
\ee
and is obtained from the $Q^{111}$ reduction by truncating a single massless vector multiplet. This amounts to setting 
\be
v_3=v_1\,,\ \ \ b_3=b_1\,,\ \ \ A^3=A^1\,. \label{M110trunc}
\ee
\subsubsection{$N^{11}$  }

The consistent truncation of M-theory on $N^{11}$ has one massless and two massive vector multiplets, along with two hypermultiplets. The scalar manifolds are
\be
\cM_v=\Blp \frac{SU(1,1)}{U(1)}\Brp^3\,,\ \ \ \cM_h= \cM_{4,2} =\frac{SO(4,2)}{SO(4)\times SO(2)}\,. 
\ee

The metric on $\cM_{2,4}$ is
\bea
ds_{4,2}^2&=&d\phi^2 +\frac{d\vphi^2}{4} +\quarter e^{-2\vphi} d\chi^2+\quarter e^{4\phi} \bslb da+ \half(\xi^0 d\txi_0-\txi_0 d\xi^0+\xi^1 d\txi_1-\txi_1 d\xi^1) \bsrb \non \\
&&+\frac{1}{8}e^{2\phi+\vphi}\blp d\xi^0+ d\xi^1 \brp^2+\frac{1}{8}e^{2\phi+\vphi}\blp d\txi_0- d\txi_1 \brp^2 \non \\
&& +\frac{1}{8}e^{2\phi-\vphi} \Bslb d\xi^0- d\xi^1 + \chi (d\txi_0- d\txi_1) \Bsrb^2 \non \\
&& +\frac{1}{8}e^{2\phi-\vphi} \Bslb d\txi_0+ d\txi_1- \chi ( d\xi^0+ d\xi^1 ) \Bsrb^2 \, , \label{SO42met}
\eea
and the special coordinate $z$ on the base is given by
\be
e^{\vphi}+i \chi= \frac{1-z}{1+z}\,,\ \ \ \ \ 
\Rightarrow\ \ \ \ \frac{1}{4}\Blp d\vphi^2 + e^{-2\vphi} d\chi^2\Brp = \frac{dz d\zbar}{(1-|z|^2)^2}\,.
\ee
This differs slightly from the special coordinate used in \cite{Cassani:2012pj},  where  the metric is taken on the upper half plane instead of the disk. The prepotential and special coordinates on $\cM_c$ are given by
\be
\cG=\frac{(Z^0)^2-(Z^1)^2}{2i}\,,\ \ \ \ Z^A=(1,z)\,.
\ee

The cubic prepotential on $\cM_v$  obtained from dimensional reduction is   the same as for  $Q^{111}$,  \eq{FQ111},  however  the models differ because
of  additional gaugings 
\be
Q_1^{\ 1}=Q_2^{\ 1}=2\,,\ \ Q_3^{\ 1}=-4\,. \label{QelecN11} 
\ee 
The duality rotation we used for the $Q^{111}$ model to make the gaugings electric would not work here since it would then make \eq{QelecN11} magnetic. However using the fact that $m^\Lam$ and $Q_\Lam^{\ 1} $ are orthogonal
\be
m^\Lam Q_{\Lam}^{\ 1}=0\,,
\ee
we can find a duality frame where all parameters are electric and $Q_{\Lam}^{\ A}$ is unchanged. Explicitly we use
\be
\cS_1= \hcR^{-1} \cS \hcR
\ee
where
\bea
\cR &=& \bpm 1 & 0& 0& 0 \\0 & c_\beta & s_\beta &0 \\ 0& -s_\beta & c_\beta &0\\0  & 0&0 & 1 \epm\bpm 1 & 0& 0& 0\\0  & 1& 0& 0 \\0 & 0& c_\al & s_\al \\ 0& 0 & -s_\al & c_\al \epm\,,\  \  \ \al=\pi/4\,,\ \ \tan\beta=\sqrt{2}\,,\non  \\
\hcR&=& \bpm \cR^{-1} & 0 \\ 0& \cR \epm\,, \\
\cS&=& \bpm A & B \\ C& D \epm\,,\ \ \ A=D= \diag(1,0,1,1)\,,\ B=-C=\diag(0,-1,0,0)\,.
\eea
The Killing vectors are then given by \eq{gaugeQ11elec} and \eq{QelecN11}.

The prepotential in this frame is rather complicated in terms of the new sections, which are in turn given as a function of the scalar fields $z^i$ by
\bea
X^\Lam&=&\frac{1}{3}(3, 2z^1-z^2-z^3+z^{123}  ,2z^2-z^1-z^3+z^{123} ,2z^3-z^1-z^2+z^{123})\,, \\
z^{123}&=& z^1 z^2 +z^2z^3 + z^3 z^1\,.
\eea

\subsubsection{Squashed $S^7$ $\sim\frac{Sp(2)}{Sp(1)}$}

This is obtained from the $N^{11}$ model by eliminating the massless vector multiplet. Explicitly, this is done by setting
\be
v_2=v_1\,,\ \ \ b_2=b_1\,, \ \ \ A^2=A^1\,.
\ee
In addition to the $\N=2$, round $S^7$ solution \eq{AdS4Sol}  this model contains in its field space  the squashed $S^7$ solution, although this vacuum has only $\cN=1$ supersymmetry. Thus flows from this solution lie outside  the ansatz employed in this work.

\subsubsection{Universal $\frac{SU(4)}{SU(3)}$ Truncation}
This model was first considered in \cite{Gauntlett:2009zw}. It contains  just one massive vector multiplet and one hypermultiplet, and  can be obtained from the $M^{111}$ truncation by setting 
\be
v_2=v_1\,,\ \ \ b_2=b_1\,,\ \ \ A^2=A^1\,.
\ee

\section{Horizon Geometries}\label{sec:hyperhorizons}

We now apply the horizon equations of Section \ref{sec:horizonEqs} to the models of Section \ref{sec:truncations}.  We find that there is a four dimensional solution space within the $Q^{111}$ model and that this governs all the other models, even though not all the other models are truncations of  $Q^{111}$. The reason is that the extra gaugings present in the $N^{11}$ and squashed $S^{7}$ model can be reinterpreted as  simple algebraic constraints on our $Q^{111}$ solution space.

 In the following, we will use the minus sign in \eq{pP1} and subsequent equations. We also recall that $\kappa =1$ refers to $AdS_2\times S^2$ and $\kappa =-1$ to $AdS_2\times \HH^2$ horizons.
 
\subsection{M-theory Interpretation}
The charges of the four-dimensional supergravity theory have a clear interpretation in the eleven-dimensional theory. This interpretation is different from how the charges lift in the theory  used in \cite{Cacciatori:2009iz}, which we now review. In the consistent truncation of M-theory on $S^7$ \cite{deWit:1984nz, Nicolai:2011cy} the $SO(8)$-vector fields lift to Kaluza-Klein metric modes in eleven-dimensions.  In the further truncation of \cite{Cvetic1999b,Duff:1999gh} only the four-dimensional Cartan subgroup of $SO(8)$ is retained, the magnetic charges of the four vector fields in \cite{Cacciatori:2009iz} lift to the Chern numbers of four $U(1)$-bundles over $\Sig_g$. One can interpret the resulting $AdS_4$ black holes as the near horizon limit of a stack of M2-branes wrapping $\Sig_g\subset X_5$,  where $X_5$ is a  praticular non-compact Calabi-Yau five-manifold, constructed as four line bundles over $\Sig_g$:
\be
\xymatrix{
\oplus_{\Lam=0}^{3}\cL_{p^\Lam}\ar[r] & X_5 \ar[d]\\
&  \Sig_g\\}
\ee
A similar description holds for wrapped D3-branes and wrapped M5-branes in the spirit of \cite{Maldacena:2000mw}. The general magnetic charge configurations have been analyzed recently for D3 branes in \cite{Benini2013a} and M5-branes in \cite{Bah:2012dg}.  Both these works have computed the field theory central charge and matched the gravitational calculation \footnote{One can also identify holographically the exact R-symmetry \cite{Szepietowski:2012tb,Karndumri:2013iqa}.}. This alone provides convincing evidence that the holographic dictionary works for general twists. There has not yet been any such computation performed from the quantum mechanics dual to the solutions of \cite{Cacciatori:2009iz},  but,  as  long as the charges are subject to appropriate quantization so as to make $X_5$ well defined, one might imagine there exist  well defined quantum mechanical duals of these solutions.

Now returning to the case at hand, the eleven-dimensional metric from which the four-dimensional theory is obtained is \cite{Cassani:2012pj}
\be
ds_{11}^2= e^{2V} \cK^{-1} ds_4^2 + e^{-V} ds_{B_6}^2+ e^{2V}(\tha+\sqrt{2}A^0)^2 \, , 
\ee
where $B_6$ is a K\"ahler-Einstein six-manifold, $\tha$ is the Sasaki fiber, $V$ is a certain combination of scalar fields (not to be confused with $V$ in \eq{metAnsatz}), $\cK=\coeff{1}{8}e^{-K}$ with $K$ the  K\"ahler potential,  and $A^0$ is the four-dimensional graviphoton\footnote{There is a factor of $\sqrt{2}$ between $A^\Lam$ here and in \cite{Cassani:2012pj}, see footnote 10 of that paper.}.
In addition, vector fields of massless vector multiplets come from the three-form potential expanded in terms of  cohomologically non-trivial two forms $\om_i$
\be
C^{(3)}\sim A^i\w \om_i\,.
\ee
The truncations discussed above come from reductions with additional,  cohomologically trivial two-forms, which give rise to the vector fields of massive vector mutliplets. This is an  important issue for our black hole solutions  since only massless vector fields carry conserved charges. \\

The solutions described in this section carry both electric and magnetic charges. The graviphoton will have magnetic charge $p^0$ given by \eq{pLamReps}, which means the eleven-dimensional geometry is really of the form
\be
AdS_2\times M_9 \, , 
\ee
where $M_9$ is a nine-manifold which can be described as a $U(1)$ fibration
\be
\xymatrix{
U(1)\ar[r] & M_9 \ar[d]\\
& B_6\times \Sig_g\\}
\ee
The electric potential $\tq^0$ will vanish from which we learn that this $U(1)$ is not fibered over $AdS_2$, or in other words the M2 branes that  wrap $\Sig_g$ do not have momentum along this $U(1)$.
In addition the charges that  lift to $G^{(4)}$ correspond to the backreaction of wrapped M2 and M5-branes on $H_2(SE_7,\ZZ)$ and $H_5(SE_7,\ZZ)$. 

We can check that the Chern number of this $U(1)$ fibration is quantized as follows. First we have
\be
\tha+\sqrt{2}A^0=d\psi +\eta +\sqrt{2} A^0
\ee
where $\psi$ has periodicity $2\pi \ell$ for some $\ell\in \RR$ and $\eta$ is a K\"ahler potential one-form on $B_6$ which satisfies $d\eta=2J$. Such a fibration over a sphere is well defined if
\be
n=\frac{\sqrt{2}}{\ell} \int \frac{dA^0}{2\pi}\in \ZZ\,.\label{nZZ}
\ee
Recalling \eq{elinv} and preempting \eq{pLamReps}, we see that 
\be
n=\frac{2\sqrt{2}}{\ell}\,p^0 =-\frac{1}{2\ell}\,.
\ee
For the $SE_7$ admitting spherical horizons used in this paper  one has 
\bea
Q^{111},N^{11}:&&\ell=\half \,, \\
M^{111}:&&\ell=\coeff{1}{4}
\eea
and \eq{nZZ} is satisfied.

\subsection{$Q^{111}$} \label{sec:Q111Horizons}
To describe the solution space of $AdS_2\times S^2$ or $AdS_2\times \HH^2$ solutions, we will exploit the fact that the gaugings \eq{gaugeQ111} are symmetric in the indices $i=1,2,3$. We can therefore express the solution in terms of invariant polynomials under the diagonal action of the symmetric group $\cS_3$\footnote{For example $\sig(v_1^2b_2)=v_1^2b_2+v_2^2b_1+v_3^2b_2+v_1^2b_3+v_2^2b_3+v_3^2b_1$ and $\sig(v_1v_2)=2(v_1v_2+v_2v_3+v_1v_3)$} 
\be
 \sig (v_{1}^{i_1}v_2^{i_2} v_3^{i_3}b_{1}^{i_1}b_2^{i_2} b_3^{i_3}) =\sum_{\sig\in S_3}v_{\sig(1)}^{i_1}v_{\sig(2)}^{i_2} v_{\sig(3)}^{i_3}b_{\sig(1)}^{i_1}b_{\sig(2)}^{i_2} b_{\sig(3)}^{i_3}\,.
 \ee
First we enforce \eq{P120}, which gives 
\be
\xi^0=0\,,\ \txi_0=0\, .
\ee
The Killing prepotentials are then given by
\be
P^3_\Lam=\sqrt{2}(4-\half  e^{2\phi }e_0,\, -e^{2\phi},\,-e^{2\phi},\,-e^{2\phi})
\ee
and the non-vanishing components of the Killing vectors by
\be
k^a_\Lam=-\sqrt{2}(e_0, 2,2,2)\,.
\ee

Solving \eq{pP} and \eq{pk} we get two constraints on the magnetic charges
\be
p^0=- \frac{1}{4\sqrt{2}}\,,\ \ \ p^1+p^2+p^3=- \frac{\sqrt{2}e_0}{16}\,. \label{pLamReps}
\ee
We find that the phase of the spinor is fixed
\be
\psi=\frac{\pi}{2} \, , 
\ee
while  \eq{tqP} and \eq{tqk} are redundant 
\be
\sig(v_1b_2)=0\,.\label{constr1}
\ee
Then from \eq{hyphor} we get
\be
\sig(v_1v_2)-\sig(b_1b_2) =e_0\,.\label{constr2}
\ee
We can of course break the symmetry and solve the equations above  for, for instance,  $(b_3,v_3)$ 
\bea
v_3&=&  \frac{v_2(e_0-2 b_1^2)-2v_1^2 v_2  +v_1(e_0-2v_1^2-2b_2^2)}{2\blp v_1^2+2v_1 v_2 + v_2^2 +(b_1+b_2)^2\brp}\,,\\
b_3&=&- \frac
{b_2(e_0+2v_1^2) +2b_1^2 b_2+b_1(e_0+2v_2^2+2b_2^2)}
{2\blp v_1^2+2v_1 v_2 + v_2^2 +(b_1+b_2)^2\brp} \,.
\eea
Using \eq{R1hor} we find the radius of $AdS_2$ to be
\bea
R_1^2&=& \frac{v_1v_2v_3}{16}\,.
\eea
The algebraic constraint \eq{Alg1} is nontrivial and can be used to solve for $q_0$ in terms of $(p^\Lam,q_{i},v_j,b_k)$. 

Using  the value of $p^0$ given in \eq{pLamReps} we can solve  \eq{phor} and \eq{qhor} and find
\bea
e^{2\phi}&=& \frac{4 (R_2^2-\kappa R_1^2)}{  R_2^2\, \sig(v_1v_2)} \, , \\
R_2^2&=&  \kappa  R_1^2 \Bslb 1-\frac{\sig(v_1v_2)^2}
{2 \hsig } \Bsrb  \, , 
 \\
&& \non \\
q_0&=&\frac{\kappa q_{0n}}{ 4\sqrt{2}\, \hsig}\, ,\\
q_{0n}&=& -\sig(v_1^3v_3 b_1^3)+\sig(v_1v_3^3b_1^2b_2) - (v_1v_2v_3)^2 \sig(b_1)-b_1b_2b_3\blp \sig(v_1^2 b_2^2)+\sig(v_1^2 b_2b_3)\brp \non \\
&&-v_1v_2v_3 \blp \sig(v_1 b_1 b_2^2) -2 \sig(v_1 b_2^2 b_3) -2 \sig(v_1^2 v_2 b_3)  \brp   \, ,  \\
&& \non \\
p^1&=& \frac{\kappa p^1_n}{ 4\sqrt{2} \hsig} \, ,    \\
p^1_n&=& 2 v_1^2v_2v_3(v_2^2+v_3^2+v_2v_3)  \, ,  \non \\
&& +v_2 v_3 (v_2^2+v_3^2) b_1^2 -2 v_1 v_2 v_3(v_2+v_3) b_2 b_3  +2(v_2^2+v_3^2)b_1^2 b_2 b_3 +2 v_1^2 b_2^2 b_3^2\non \\
&&-\Bslb \blp 2 v_1 v_3^2(v_2+v_3) b_1 b_3 +(-v_1^2 v_2+2 v_1 v_2v_3  + (2 v_1+v_2) v_3^2)v_3 b_2^2 \brp + (2\lra 3) \Bsrb \non \\
&&+ \Bslb 2 v_3^2 b_1b_2^2 b_3 + (v_1^2+v_3^2) b_2^3 b_3  + (2\lra 3) \Bsrb  \, ,    \\
&& \non \\
q_1&=& \frac{\kappa q_{1n}}{4\sqrt{2}\, \hsig}   \, ,   \\
q_{1n}&=& -v_1v_2v_3 \sig(v_1) b_1  -\Bslb   v_1^2b_2  \sig(v_1v_2) +(2\lra 3)\Bsrb \non \\
&&+2 v_1^2 b_1b_2b_3 +\Bslb v_2^2b_1^3 + 2 v_3^2 b_1^2 b_2 + (v_1^2+v_3^2) b_1 b_2^2 +(2\lra 3)\Bsrb  \, ,  
\eea
where
\be
\hsig= v_1v_2v_3 \sig(v_1) - \sig(v_1^2 b_2^2)-\sig( v_1^2 b_2b_2)\,.
\ee
The  charges $(p^2,p^2,q_2,q_3)$ are related to $(p^1,q_1)$ by symmetry of the $i=1,2,3$ indices.  

The general solution space has been parameterized by $(v_i,b_j)$ subject to the two constraints \eq{constr1} and \eq{constr2} leaving a four dimensional space. From these formula, one can easily establish numerically  regions where the horizon geometry is regular. A key step omitted here is to invert these formulae and express the scalars $(b_i,v_j)$ in terms of the charges $(p^\Lam,q_\Lam)$. This would allow one to express the entropy and the effective $AdS_2$ radius in term of the charges \cite{wip}.

\subsubsection{A $Q^{111}$ simplification} \label{sec:Q111Simp}

The space of solutions in the $Q^{111}$ model simplifies considerably if one enforces a certain symmetry
\be
p^1=p^2\,,\ \ \ q_1=-q_2\,.\label{Q111Simp}
\ee
One then finds a two-dimensional space of solutions part of which was found in \cite{Donos:2008ug, Donos2012d} 
\bea
v_2&=&v_1\,,\ \ \ b_3=0\,,\ \ \ b_2=-b_1\\
b_1&=& \eps_1 \sqrt{\frac{ e_0-2v_1^2-4v_1v_3}{2}} \\
e^{2\phi} &=&\frac{4(v_1+2v_3)}{ v_1(e_0+6v_3^2)}\\
R_1&=&\frac{v_1 \sqrt{v_3}}{4} \\
R_2^2&=& R_1^2 \frac{\kappa ( e_0+6v_3^2)}{\blp e_0-2(v_1^2+4 v_1 v_3 + v_3^2)\brp} \\
q_0&=& 0 \\
q_1&=&-\kappa \eps_1 \frac{(e_0-4 v_1 v_3-2v_3^2 )\sqrt{e_0-4 v_1v_3 - 2v_1^2}}{8\blp e_0-2(v_1^2+4 v_1 v_3 + v_3^2)\brp}  \label{q1Q111Simp}\\
q_3&=& 0\\
p^0&=&- \frac{1}{4\sqrt{2}} \\
 p^1&=&- \frac{ v_1 v_3 \blp e_0+2 v_3^2- 2 v_1 v_3 \brp}{4\sqrt{2}\blp e_0-2(v_1^2+4 v_1 v_3 + v_3^2)\brp}  \label{p1Q111Simp} \\
 p^3&=& \frac{\sqrt{2}e_0}{16}-2p^1 \,,
\eea
where $\eps_1=\pm$ is a choice. 
One cannot analytically invert \eq{q1Q111Simp} and \eq{p1Q111Simp} to give $(v_1,v_3)$ in terms of $(p^1,q_1)$ but one can numerically map the space of charges for which regular solutions exist.

\subsection{$M^{111}$} \label{sec:M110Solutions}

The truncation to the $M^{111}$ model \eq{M110trunc} does not respect the simplification \eq{Q111Simp}. The general solution space is two-dimensional
\bea
b_3&=&b_1\,,\ \ \ v_3=v_1\,,\ \ \ p^3=p^1\,,\ \ \ q_3=q_1\,, \\
b_1&=& \eps_2 \sqrt{\frac{v_1(e_0-2v_1(v_1+2v_2))}{2(v_1+2v_2)}}  \, ,  \\
b_2&=&-\frac{(v_1+v_2)b_1}{v_1}  \, ,  \\
e^{2\phi}&=& \frac{4(v_1+2v_2)^2}{2v_1^4+8 v_1^3v_2+(3e_0+8v_1^2)v_2^2}  \, ,  \\
R_1&=& \frac{v_1 \sqrt{v_2}}{4}  \, ,  \\
R_2^2 &=& \kappa R_1^2 \frac{ (2v_1^4 + 8v_1^3v_2 +(3 e_0+8 v_1^2)v_2^2)}{v_2(3 e_0 v_2-4v_1(v_1+2v_2)^2)}\, ,
\eea
\bea
p^0&=&-\frac{1}{4\sqrt{2}}\,, \label{p0M110}    \\
p^2&=&\frac{e_0}{8\sqrt{2}}-2p^1\, , \\
p^1&=&- \frac{e_0}{8\sqrt{2}}\frac
{2v_1^4-3e_0  v_2  (v_1+v_2)+12 v_1^2 v_2 (v_1+2v_2)+16 v_1 v_2^3}
{(v_1+2v_2)(3 e_0 v_2-4v_1(v_1+2v_2)^2)}   \, ,  \\
q_0&=&- \frac{\kappa \eps_2 }{16} \sqrt{\frac{v_1(e_0-2v_1(v_1+2v_2))}{(v_1+2v_2)^3}} \non \\
&& \cdot \frac{8v_1^6 -v_2 (v_1+v_2)(  3e_0^2 +4e_0 v_1^2 - 48 v_1^4)+ 48 v_1^4 v_2^2+ 8v_1 v_2^3(e_0+8 v_1^2) }{3 e_0 v_2 -4v_1(v_1+2v_2)^2}   \, ,  \non \\ && \\
q_1&=&-\frac{\kappa \eps_2}{8} \sqrt{\frac{v_1(e_0-2v_1(v_1+2 v_2))}{(v_1+2v_2)}}
\frac{3 e_0 v_2 -2v_1(v_1+2v_2)^2}{3 e_0 v_2 -4v_1(v_1+2v_2)^2}  \, ,  \\
q_2&=& -\frac{\kappa \eps_2}{8} \sqrt{\frac{(e_0-2v_1(v_1+2 v_2))}{v_1(v_1+2v_2)}}
 \frac{4 v_1^4 +v_2(16 v_1^2-3e_0)(v_2+v_1)}{3 e_0 v_2 -4v_1(v_1+2v_2)^2}   \, ,  \label{q3M110}
\eea
where $\eps_2$ is a choice of sign.

\subsection{$N^{11}$}

In setting $P^1_\Lam=P^2_\Lam=0 $ we get
\be
\xi^A=\txi_A=0\,,\ \ \  z^1=\zbar^1=0 \, , 
\ee
and so the only remaining hyper-scalars are $(\phi,a)$. With this simplification
the  Killing prepotentials are the same as for $Q^{111}$
\be
P^3_\Lam=\sqrt{2}(4-\half  e^{2\phi }e_0,\, -e^{2\phi},\,-e^{2\phi},\,-e^{2\phi}) \, ,
\ee
while the Killing vectors have an additional component in the $\xi^1$-direction:
\bea
k^a_\Lam&=&-\sqrt{2}(e_0, 2,2,2)\,, \\
k^{\xi^1}_{\Lam}&=& \sqrt{2}(0,-2,-2,4)\,.
\eea

From this one can deduce that the spectrum of horizon solutions will be obtained from that of $Q^{111}$ by imposing two additional constraints
\bea
p^\Lam k^{\xi^1}_{\Lam}&=& 0\,, \\
\tq^\Lam k^{\xi^1}_{\Lam}&=& 0\,,
\eea
which amount to 
\bea
p^3&=& \half (p^1+p^2 )\,, \label{N11constraint1}\\
v_3&=& \half (v_1+v_2 )\,.\label{N11constraint2}
\eea

One can then deduce that the $AdS_2\times \Sig_g$ solution space in the $N^{11}$ model is a two-dimensional restriction of the four dimensional space from the $Q^{111}$ model. While \eq{N11constraint2} can easily be performed on the general solution space, it is somewhat more difficult to enforce \eq{N11constraint1} since the charges are given in terms of the scalars. We can display explicitly a one-dimensional subspace of the $N^{11}$ family by further setting $v_3=v_1$:
\bea
v_1&=& \sqrt{\frac{e_0+6(\sqrt{3}-2)b_2^2}{6}} \, , \\
b_1&=& -\frac{b_2}{2}\Blp \sqrt{3(7-4\sqrt{3})}+1\Brp   \, , \\
b_3&=& -\frac{b_2}{2}\Blp -\sqrt{3(7-4\sqrt{3})}+1\Brp   \, , \\
R_1^2&=& \frac{1}{16}  v_1^{3/2}   \, , \\
R_2^2&=& -\frac{\kappa}{48 \sqrt{6}}\frac{(e_0+3(\sqrt{3}-2)b_2^2)(e_0+6(\sqrt{3}-2)b_2^2)^{3/2}}{(e_0+12 (\sqrt{3}-2)b_2^2)}  \, , \\
e^{2\phi}&=& \frac{6}{e_0+3(\sqrt{3}-2)b_2^2}  \, , \\
p^1&=& \frac{e_0(e_0-6 b_2^2)}{24\sqrt{2}(e_0+12 (\sqrt{3}-2)b_2^2)}  \, , \\
p^2&=& \frac{e_0(e_0-6(4\sqrt{3}-7) b_2^2)}{24\sqrt{2}(e_0+12 (\sqrt{3}-2)b_2^2)}  \, , \\
p^3&=& \frac{e_0}{24\sqrt{2}}  \, , \\
q_0&=& -\frac{\kappa b_2^2\Blp (5-3\sqrt{3})e_0^2+ 9(11\sqrt{3}-19)e_0 b_2^2 +18(71-41\sqrt{3})b_2^4\Brp}
{2\sqrt{2}(e_0+6(\sqrt{3}-2)b_2^2)(e_0+12 (\sqrt{3}-2)b_2^2)}  \, , \\
q_1&=& - \frac{3\kappa b_2^2 (7-4\sqrt{3})}{2\sqrt{2}(e_0+12 (\sqrt{3}-2)b_2^2)}  \, , \\
q_2&=&- \frac{3\kappa b_2^2 (-2+\sqrt{3})}{2\sqrt{2}(e_0+12 (\sqrt{3}-2)b_2^2)}  \, , \\
q_3&=& -\frac{3\kappa b_2^2 (-5+3\sqrt{3})}{2\sqrt{2}(e_0+12 (\sqrt{3}-2)b_2^2)} \, . 
\eea

\subsection{$\frac{Sp(2)}{Sp(1)}$}

The truncation of M-theory on $\frac{Sp(2)}{Sp(1)}$ is obtained from the $N^{11}$ truncation
by removing a massless vector multiplet. Explcitly, this is done by setting
\be
v_2=v_1\,,\ \ \ b_2=b_1\,, \ \ \ A^2=A^1\,.
\ee
Alternatively one can set 
\be
p^2=p^1\,,\ \ v_2=v_1
\ee
on the two-dimensional  $M^{111}$ solution space of Section \ref{sec:M110Solutions}.  This leaves a unique solution, the universal solution of $\frac{SU(4)}{SU(3)}$ we next describe.

\subsection{$\frac{SU(4)}{SU(3)}$}\label{sec:SU4SU3Sols}
This solution is unique and requires $\kappa=-1$.  Therefore it only exists for hyperbolic horizons:
\bea
v_1&=& \sqrt{\frac{e_0}{6}}\,, \\
b_1&=& 0\,, \\
R_1&=& \frac{1}{4}\Blp \frac{e_0}{6}\Brp^{3/4}\,, \\
R_2&=& \frac{1}{2\sqrt{2}}\Blp \frac{e_0}{6}\Brp^{3/4}\,.
\eea
It is connected to the central $AdS_4$ vacuum by a flow with constant scalars, which is known analytically \cite{Caldarelli1999} .

\section{Black Hole solutions: numerical analysis}\label{numerical}

Spherically symmetric, asymptotically $AdS$ static black holes  can be seen as  solutions interpolating  between  $AdS_4$ and $AdS_2\times S^2$.
We have seen that  $AdS_2\times S^2$  vacua are quite generic
in the consistent truncations of M-theory on Sasaki-Einstein spaces and we may expect that they arise as horizons of  static black holes.
In this section we will show that this is the case in various examples  and we expect that this is true in general.

The system of BPS equations (\ref{pP1}) - (\ref{psiEq}) can be consistently truncated to the locus 
\begin{equation}\label{hyperlocus}
\xi^A =0\, , \qquad \tilde\xi_A=0 \, ;
\end{equation}
this condition   is satisfied at the fixed points and  enforces (\ref{P120}) along the flow. The only running hyperscalar is the dilaton $\phi$.  The solutions of (\ref{pP1}) - (\ref{psiEq})  will have  a non trivial profile for the dilaton, all  the scalar fields in the vector multiplets, the gauge fields and  the phase of the spinor.  This makes it hard to solve the equations  analytically. We will find asymptotic solutions near   $AdS_4$ and $AdS_2\times S^2$ by expanding the
equations in series and  will  find an interpolating solution numerically. The problem simplifies when symmetries allow to set all the massive gauge fields and the phase of the spinor to zero. A solution of this form can be found in the model corresponding to the
truncation on   $Q^{111}$. The corresponding solution is discussed in Section  \ref{numericalQ111} and it  corresponds to the class of solutions found  in eleven dimensions in \cite{Donos2012d}. The general case is more complicated. The $M^{111}$ solution discussed in Section \ref{numericalM110} is an example of the general case, with most of the fields turned on.

\subsection{Black Hole solutions in $Q^{111}$}\label{numericalQ111}

We now construct a black hole interpolating between the $AdS_4 \times Q^{111}$ vacuum and the horizon solutions discussed
in Section \ref {sec:Q111Simp} with
\be
p^1=p^2\,,\ \ \ q_1=-q_2\,.\label{Q111Simp2}
\ee
The solution should correspond to the M-theory one found  in \cite{Donos2012d}. Due to the high degree of symmetry of the model,  we can  truncate the set of fields appearing 
in the solution and consistently set
\begin{equation}
v_2=v_1\,,\ \ \ b_3=0\,,\ \ \ b_2=-b_1
\end{equation}
along the flow. This restriction is compatible with the following simplification on the gauge fields 
\begin{equation}
\tilde q_2(r)=-\tilde q_1(r)\,,\ \ \ \tilde q_0(r)=0\,,\ \ \ \tilde q_3(r)=0\, .
\end{equation}
It follows that
\begin{equation}\label{simpcond}
k^a_\Lambda \, \tilde q^\Lambda =0\, , \\\ \qquad  P^3_\Lambda\,  \tilde q^\Lambda =0 
\end{equation}
for all $r$. The latter conditions lead to several interesting simplifications. $k^a_\Lambda \, \tilde q^\Lambda =0$ implies that the right hand side of Maxwell equations (\ref{Max1}) vanishes and  no massive vector field is turned on. Maxwell equations then reduce  to conservation of the invariant electric charges $q_\Lambda$, and we can use the definition (\ref{maginv}) to find an algebraic expression for $\tilde q_\Lambda$ in terms of the scalar fields. Moreover, the condition $P^3_\Lambda\,  \tilde q^\Lambda =0$ implies that the phase $\psi$ of the spinor is constant along the flow. Indeed, with our choice of fields,  $A_r=0$ and the equation (\ref{psiEq}) reduces to $\psi^\prime =0$. The full set of BPS equations reduces to six first order equations for the six quantities
\begin{equation}
\{ U,V,v_1,v_3,b_1,\phi \} \, .
\end{equation}

\vspace{0.2cm}

For simplicity, we  study the interpolating solution corresponding to the horizon solution  in Section \ref {sec:Q111Simp} with $v_1=v_3$. This restriction leaves a family of $AdS_2\times S^2$ solutions which can be parameterized by the value of $v_1$ or, equivalently, by the magnetic charge $p^1$. We perform our numerical analysis for the model with
\be
e^{-2\phi} = \frac{11}{6\sqrt{2}}\, , \qquad  v_1 = v_3 = \frac{1}{2^{1/4}}\, , \qquad b_1 = - \frac{\sqrt{5}}{2^{1/4}} 
\ee
and electric and magnetic charges
\be 
p^1=- \frac12 \, , \qquad q_1=  \frac{5\sqrt{5}}{8 \, 2^{3/4}}\, .
\ee
We  fixed $e_0=8 \sqrt{2}$. The values of the  scalar fields at the $AdS_4$ point are given in (\ref{AdS4Sol}).  \\

 It is convenient to define a new radial coordinate by $dt= e^{-U} dr$. $t$  runs from $+\infty$ at the $AdS_4$ vacuum to $-\infty$ at the horizon. It is also convenient
 to  re-define some of the scalar fields
\begin{equation}
 v_i(t) = v_i^{AdS} e^{e_i(t)}\, , \qquad  \phi(t)=\phi_{AdS} -\frac12 \rho(t) \, ,
 \end{equation}
  such that they vanish at the $AdS_4$ point. The metric functions will be also re-defined 
  \begin{equation}
  U(t) =u(t)+\log(R_{AdS})\, , \qquad V(t)=v(t) 
  \end{equation}
   with $u(t)=t,v(t)=2t$ at the $AdS_4$ vacuum.
The BPS equations read

\begin{eqnarray}
u'&=&
 e^{-e_1-\frac{e_3}{2}}  - \frac{3}{4} e^{-e_1-\frac{e_3}{2}-\rho} +\frac{1}{4} e^{e_1-\frac{e_3}{2}-\rho}
 +\frac{1}{2} e^{\frac{e_3}{2}-\rho} +\frac{3}{8} e^{-\frac{e_3}{2}+2 u-2 v} -\frac{3}{4} e^{-e_1+\frac{e_3}{2}+2 u-2 v}\nonumber\\
&&-\frac{1}{8} e^{e_1+\frac{e_3}{2}+2 u-2 v} - \frac{15 \sqrt{5} e^{-e_1+\frac{e_3}{2}+2 u-2 v} b_1}{32\ 2^{3/4}}+\frac{3 e^{-e_1-\frac{e_3}{2}-\rho} b_1^2}{16 \sqrt{2}}-\frac{3 e^{-e_1+\frac{e_3}{2}+2 u-2 v} b_1^2}{32 \sqrt{2}}\, , \nonumber\\
v'&=& 2 e^{-e_1-\frac{e_3}{2}}-\frac{3}{2} e^{-e_1-\frac{e_3}{2}-\rho}+\frac{1}{2} e^{e_1-\frac{e_3}{2}-\rho}+e^{\frac{e_3}{2}-\rho}+\frac{3 e^{-e_1-\frac{e_3}{2}-\rho} b_1^2}{8 \sqrt{2}}\, ,\nonumber\\
e'_1&=& 
2 e^{-e_1-\frac{e_3}{2}} -\frac{3}{2} e^{-e_1-\frac{e_3}{2}-\rho} -\frac{1}{2} e^{e_1-\frac{e_3}{2}-\rho} +\frac{3}{2} e^{-e_1+\frac{e_3}{2}+2 u-2 v}
-\frac{1}{4} e^{e_1+\frac{e_3}{2}+2 u-2 v}\nonumber \\
&&+\frac{15 \sqrt{5} e^{-e_1+\frac{e_3}{2}+2 u-2 v} b_1}{16\ 2^{3/4}} +\frac{3 e^{-e_1-\frac{e_3}{2}-\rho} b_1^2}{8 \sqrt{2}}
+\frac{3 e^{-e_1+\frac{e_3}{2}+2 u-2 v} b_1^2}{16 \sqrt{2}} \, ,\\
e'_3&=& 2 e^{-e_1-\frac{e_3}{2}} -\frac{3}{2} e^{-e_1-\frac{e_3}{2}-\rho} +\frac{1}{2} e^{e_1-\frac{e_3}{2}-\rho} -e^{\frac{e_3}{2}-\rho}
-\frac{3}{4} e^{-\frac{e_3}{2}+2 u-2 v}  -\frac{3}{2} e^{-e_1+\frac{e_3}{2}+2 u-2 v}\nonumber\\
&&-\frac{1}{4} e^{e_1+\frac{e_3}{2}+2 u-2 v} -\frac{15 \sqrt{5} e^{-e_1+\frac{e_3}{2}+2 u-2 v} b_1}{16\ 2^{3/4}}
+\frac{3 e^{-e_1-\frac{e_3}{2}-\rho} b_1^2}{8 \sqrt{2}}  -\frac{3 e^{-e_1+\frac{e_3}{2}+2 u-2 v} b_1^2}{16 \sqrt{2}} \, ,\nonumber\\
b'_1&=&  - \frac{5 \sqrt{5} e^{e_1+\frac{e_3}{2}+2 u-2 v}}{4\ 2^{1/4}} -e^{e_1-\frac{e_3}{2}-\rho} b_1- \frac{1}{2} e^{e_1+\frac{e_3}{2}+2 u-2 v} b_1 \, ,\nonumber\\
\rho'&=&
- 3 e^{-e_1-\frac{e_3}{2}-\rho}+e^{e_1-\frac{e_3}{2}-\rho}+2 e^{\frac{e_3}{2}-\rho}+\frac{3 e^{-e_1-\frac{e_3}{2}-\rho} b_1^2}{4 \sqrt{2}} \, .\nonumber
\end{eqnarray}

This set of equations has two obvious symmetries. Given a solution, we can generate other ones by 
\be
u(t)\rightarrow u(t) + d_1 \, ,  \qquad v(t)\rightarrow v(t) + d_1  \, , \label{simm1}
\ee
or by translating all fields $\phi_i$ in the solution
\be
\phi_(t) \rightarrow \phi_i(t- d_2)  \, , \label{simm2}
\ee
where $d_1$ and $d_2$ are arbitrary constants. \\

We can expand the equations  near the $AdS_4$ UV point. We should stress again  that $AdS_4$ is not strictly a solution 
due to the presence of a magnetic charge at infinity.
However, the metric functions $u$ and $v$ approach the $AdS_4$ value and, for large $t$, the linearized equations of motion for the scalar fields are not affected by the magnetic charge, so that we can use much of the intuition  from the AdS/CFT correspondence.
The  spectrum of the consistent truncation around the $AdS_4$ vacuum  in absence of charges have been analyzed in details in \cite{Cassani:2012pj}. It consists of two massless and one massive vector multiplet (see Table 1). By expanding the BPS equations for large $t$
we find that there exists a family of asymptotically (magnetic) $AdS$ solutions depending on three parameters, corresponding to two
operators of dimension $\Delta=1$ and an operator of dimension $\Delta=4$. The asymptotic expansion of the solution is  

\begin{eqnarray}\label{expUVQ111}
u(t) &=& t+ \frac{1}{64} e^{-2 t} \left(16-6 \epsilon_1^2-3 \sqrt{2} \beta_1^2\right)+ \cdots \nonumber \\
v(t) &=& 2 t -\frac{3}{32} e^{-2 t} \left(2 \epsilon_1^2+\sqrt{2} \beta_1^2\right) +\cdots \nonumber \\
e_1(t)&=& 
-\frac{1}{2} e^{-t} \epsilon_1+\frac{1}{80} e^{-2 t} \left(-100-4 \epsilon_1^2-3 \sqrt{2} \beta_1^2\right) +\cdots \non \\
&& +\frac{1}{140} e^{-4 t} \left(140 \epsilon_4+ 
\left(-\frac{375}{8}+ \cdots \right) t\right) +\cdots \nonumber\\ 
e_3(t)&=&
e^{-t} \epsilon_1+\frac{1}{80} e^{-2 t} \left(200-34 \epsilon_1^2-3 \sqrt{2} \beta_1^2\right) +\cdots \non \\
&&+e^{-4 t} \frac{1}{448}  \left(1785 + 448 \epsilon_4 - 150 t+\cdots \right ) +\cdots \nonumber\\
b_1(t)&=& e^{-t} \beta_1+e^{-2 t} \left(\frac{5 \sqrt{5}}{4\ 2^{1/4}}-\epsilon_1 \beta_1\right)+ \cdots \\
\rho(t) &=& \frac{3}{40} e^{-2 t} \left(2 \epsilon_1^2-\sqrt{2} \beta_1^2\right)+\cdots +\frac{1}{17920}e^{-4 t} \left(-67575-26880 \epsilon_4
+9000 t  +\cdots \right) +\cdots \, .\nonumber 
\end{eqnarray}
where the dots refer to exponentially suppressed terms in the expansion in $e^{-t}$ or to terms at least quadratic in the parameters $(\epsilon_1,\epsilon_4,\beta_1)$. We also set two arbitrary constant terms appearing in the expansion of $u(t)$ and $v(t)$ to zero for notational simplicity; they can be restored by applying the
transformations (\ref{simm1}) and (\ref{simm2}).  The constants $\epsilon_1$ and $\beta_1$ correspond to scalar modes of dimension $\Delta=1$ in the two different massless vector multiplets (cfr  Table 7 of \cite{Cassani:2012pj}). The constant $\epsilon_4$ corresponds to a scalar mode with $\Delta=4$ belonging to the massive vector multiplet. A term  $t e^{-4t}$ shows up at the same order as $\epsilon_4$ and it  is required for consistency. Notice that, although $e_1=e_3$ both at the UV and IR, the mode $e_1-e_3$ must be turned on along the flow.

In the IR, $AdS_2\times S^2$ is an exact solution of the BPS system. The relation between the two radial coordinates is $r-r_0\sim e^{ a t }$ with  $a= 8 \, 2^{1/4} /3^{1/4}$, where $r_0$ is the position of the horizon.  By linearizing the BPS equations around  $AdS_2\times S^2$ we find three normalizable modes
with behavior $e^{a \Delta t}$ with $\Delta = 0$, $\Delta=1$ and $\Delta=1.37$. The IR expansion is obtained as a double series in  $e^{a  t}$ and $e^{1.37 a  t}$ 
\begin{eqnarray}
 \{u(t),v(t),e_1(t),e_3(t),b_1(t),\rho(t) \} =&& \hskip -0.6truecm \{  1.49 + a \, t , 0.85 + a\, t , -0.49, -0.49, -1.88, -0.37 \}  \nonumber \\
&&  \hskip -0.8truecm +\{1, 1, 0, 0, 0, 0 \} c_1  + \{-1.42, -0.53, 0.76, 
  0.53, -0.09,   1\} \,  c_2 \, e^{a  t}  \nonumber \\
  &&   \hskip -0.8truecm  +\{0.11, 0.11, 0.07, 
  0.93, -0.54, 1\} \, c_3 \, e^{1.37 a  t}\nonumber \\ &&  \hskip -0.8truecm +\sum_{p,q} {\vec d}^{p,q} c_2^p c_3^q e^{(p + 1.37 q) a  t}\, ,
  \end{eqnarray}
 where the numbers ${\vec d}^{p,q}$ can be determined numerically at any given order. The two symmetries (\ref{simm1}) and (\ref{simm2}) are manifest in this expression and correspond to combinations of a shift in $c_1$ and  suitable rescalings of $c_2$ and $c_3$.

\begin{figure}[h]
\begin{minipage}[t]{\linewidth}
~~~\begin{minipage}[t]{0.5\linewidth}
\vspace{0pt}
\centering
\includegraphics[scale=.6]{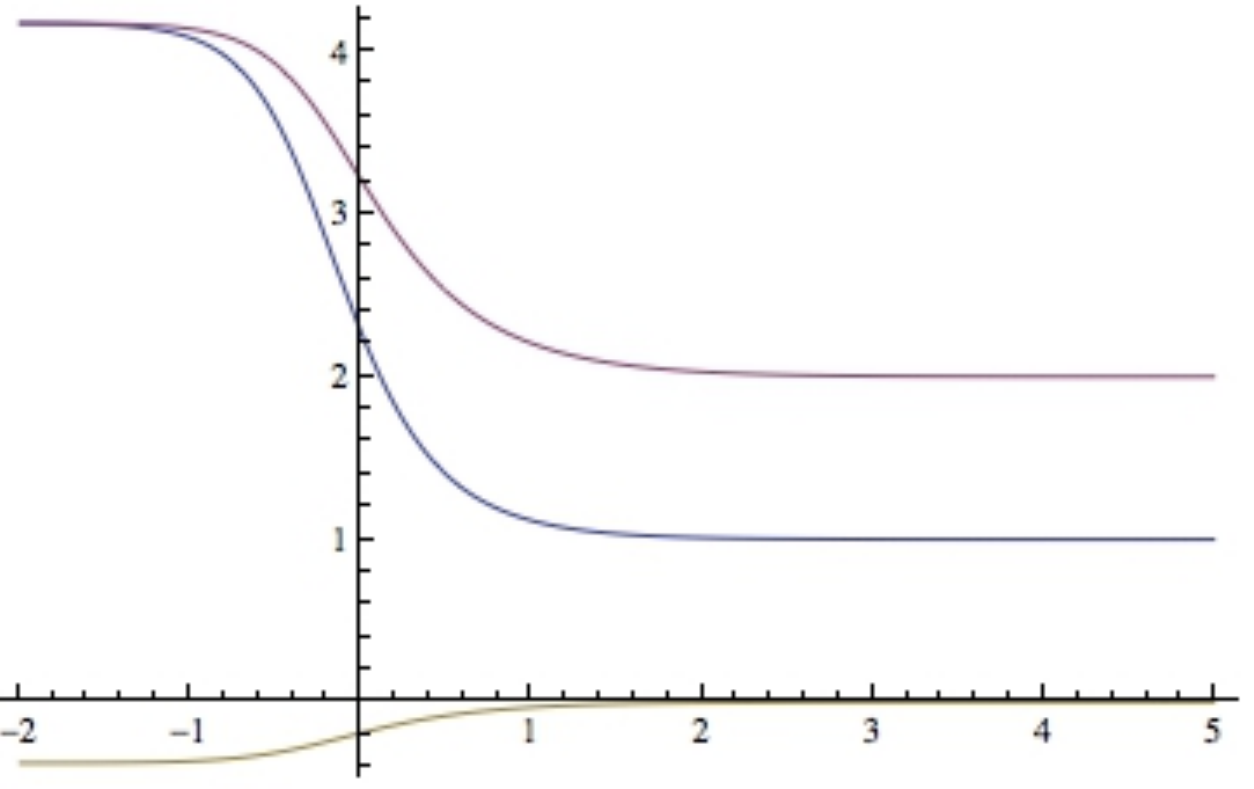}
\end{minipage}%
\hspace{.5cm}
\begin{minipage}[t]{0.5\linewidth}
\vspace{0pt}
\centering
\includegraphics[scale=.6]{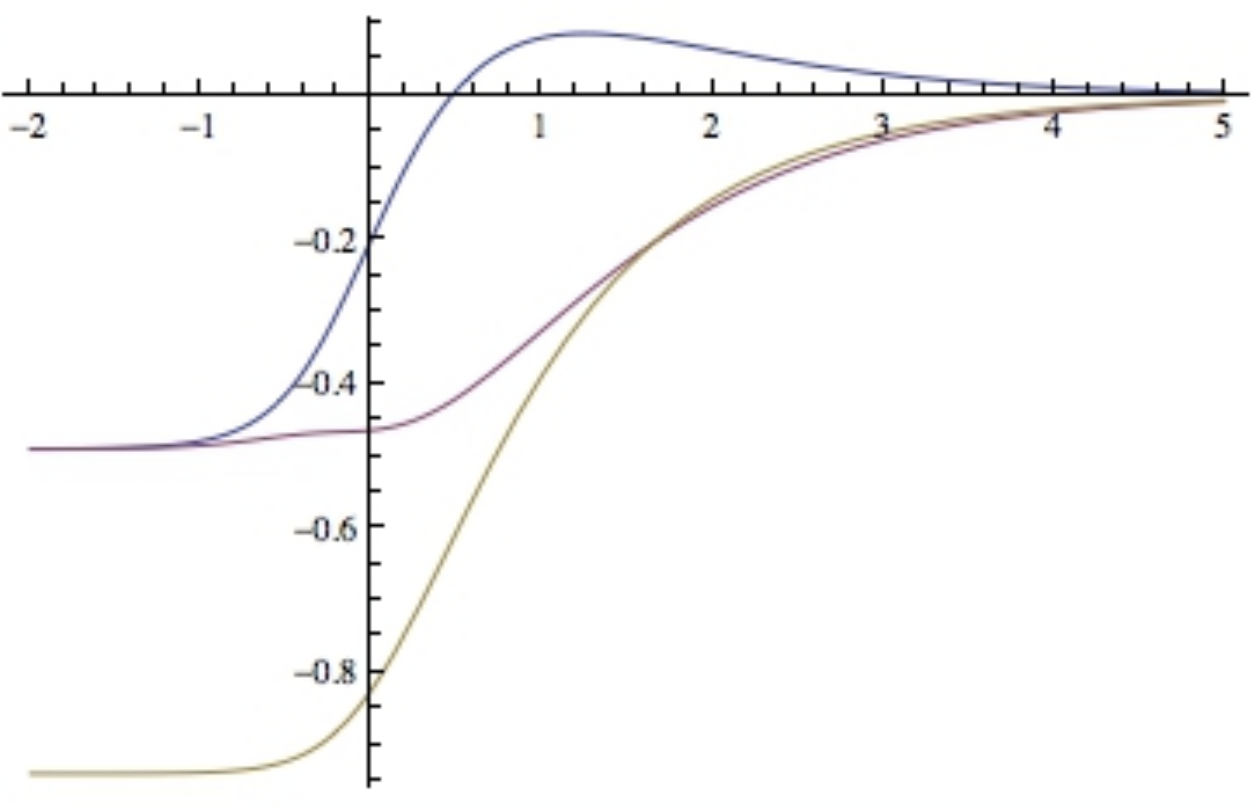}
\end{minipage}\\[1em]
\begin{minipage}[t]{0.5\linewidth}
\vspace{0pt}
\centering
\end{minipage}%
\vspace{1em}
\caption {Plots of $u',v'$ and $\rho$ on the left and of $e_1,e_2$ and $b_1/2$ on the right corresponding to 
the IR parameters $c_1=-1.208,c_2=0.989,c_3=-0.974$ and the UV parameters $\beta_1=-2.08,\epsilon_1=-1.325, \epsilon_4=5$. }
\label{fig:flowQ111}
\end{minipage}
\end{figure}

 With a total number of six parameters for six equations we expect that the given IR and UV expansions can be matched at some point in  the middle, since the equations are first order and the number of fields is equal to the number of parameters.  There will be precisely one solution with the UV and IR  asymptotics given above;  the general
 solution will be obtained by applying the
transformations (\ref{simm1}) and (\ref{simm2}).  We have numerically solved the system of BPS equation and tuned the parameters in order to find an interpolating solution. The result is shown in Figure \ref{fig:flowQ111}. \\
 
We would like to stress that the asymptotic expansions of the solutions contain integer powers of $r$ (and logs) in the UV ($AdS_4$) and irrational powers depending on the charges in the IR ($AdS_2\times S^2$). This suggests that it would be hard
to find analytic solutions of the system of BPS equations  (\ref{pP1}) - (\ref{psiEq}) with running hypermultiplets. By contrast, the static $AdS_4$ black holes in theories without hypermultiplets  \cite{Cacciatori:2009iz} depends only on  rational functions of $r$ which made it  possible to find an explicit analytic solution. 
\subsection{Black Hole solutions in $M^{111}$}\label{numericalM110}

Whenever we cannot enforce any symmetry on the flow, things are much harder. This is the case of the interpolating solutions for $M^{111}$ which we now discuss.
The solution can be also embedded in the $Q^{111}$ model and it is a general prototype of the generic interpolating solution between $AdS_4$ and the  horizons solutions discussed in Section \ref{sec:hyperhorizons}. \\

Let us consider an interpolating solution corresponding to the horizon discussed in Section  \ref{sec:M110Solutions}. The conditions (\ref{simpcond}) cannot be imposed along the flow. As a consequence, the phase of the spinor will run and
a massive gauge field will be turned on. Moreover, the IR conditions $b_2=-2 b_1$ and $\tilde q_0=\tilde q_3=0, \tilde q_2=-\tilde q_1$ do not hold for finite $r$ and all gauge and vector scalar fields are turned on. The only simplification comes from the fact that on the locus (\ref{hyperlocus})  the right hand side of Maxwell equations (\ref{Max1}) is proportional to $k^a_\Lambda$. For $M^{111}$, $k^a_1=k^a_2$ and we still have two conserved electric charges
\begin{equation}
( q_1 -q_2 )' =0\, , \qquad ( k^a_1 q_0 -k^a_0 q_1)' =0 \, .
\end{equation}
In other words, two Maxwell equations can be reduced to  first order constraints while the third remains second order. It is convenient to transform the latter equation into a pair of first order constraints.
This can be done by introducing $q_0$ as a new independent field and by using one component of  Maxwell equations  and the definition (\ref{maginv}) of $q_\Lambda$ as a set of
four first order equations for ($\tilde q_0,\tilde q_1,\tilde q_2, q_0$). The set of BPS and Maxwell equations consists of twelve first order equations for twelve variables
\begin{equation}
\{ U,V,v_1,v_2,b_1,b_2,\phi, \psi ,\tilde q_0,\tilde q_1,\tilde q_2, q_0   \} \, .
\end{equation}

A major simplification arises if we integrate out the gauge fields using (\ref{qLr}). The system collapses to a set of eight first order equations for eight unknowns. The resulting set of equations have singular denominators and
it is convenient to keep the extra field $q_0$ and  study a system of nine first order equations for
\begin{equation}
\{ U,V,v_1,v_2,b_1,b_2,\phi, \psi , q_0   \} \, .
\end{equation}
The final system has an integral of motion which would allow to eliminate algebraically $q_0$ in terms of the other fields. \\

The system of BPS equations is too long to be reported here but it can be studied numerically and by power series near the UV and the IR.  We will study the flow to the one-parameter family of horizon solutions with $v_1=v_2$ and $b_2=-2 b_1$. 
These horizons can be parametrized by the value of $v_1$ or, equivalently, by the magnetic charge $p^2$. 
We perform our numerical analysis for the model with
\be
e^{-2\phi} = \frac{5}{\sqrt{2}}\, , \qquad  v_1 = v_2 = 2^{1/4}\, , \qquad b_1 = \sqrt{3}\,  2^{1/4} 
\ee
and electric and magnetic charges
\be 
p^2=- 2 \, , \qquad q_2=  \frac{3\sqrt{3}}{4 \, 2^{1/4}}\, .
\ee
We  fixed $e_0=24 \sqrt{2}$. The values of the  scalar fields at the $AdS_4$ point are given in (\ref{AdS4Sol}). As in the previous section,  it is also convenient  to define a new radial coordinate by $dt= e^{-U} dr$ and to re-define some of the scalar fields and metric functions 
\begin{equation}
 v_i(t) = v_i^{AdS} e^{e_i(t)}\, , \,\,\,  \phi(t)=\phi_{AdS} -\frac12 \rho(t)\, , \,\,\,  U(t) =u(t)+\log(R_{AdS})\, , \,\,\, V(t)=v(t) \, .
 \end{equation}

In absence of charges, the  spectrum of the consistent truncation around the $AdS_4$ vacuum  consists of one massless and one massive vector multiplet  \cite{Cassani:2012pj} (see Table 1).
By expanding the BPS equations for large $t$ we find that there exists a family of asymptotically (magnetic) $AdS$ solutions depending on three parameters corresponding to 
operators of dimension $\Delta=1$,  $\Delta=4$ and $\Delta=5$. The asymptotic expansion of the solution is

\begin{eqnarray}
u(t) &=& t -\frac{1}{64} e^{-2 t} \left(-16+24 \epsilon_1^2+3 \sqrt{2} \beta_1^2\right)+\cdots \nonumber\\
v(t) &=& 2 t  -\frac{3}{32} e^{-2 t} \left(8 \epsilon_1^2+\sqrt{2} \beta_1^2\right)+\cdots \nonumber\\
e_1(t) &=& e^{-t} \epsilon_1-\frac{1}{80} e^{-2 t} \left(-60+16 \epsilon_1^2+3 \sqrt{2} \beta_1^2\right) +\cdots \non \\
&& -\frac{e^{-4 t} (1317+7168 \rho_4+864 t +\cdots )}{10752} +\cdots \nonumber\\
e_2(t) &=& 
-2 e^{-t} \epsilon_1-\frac{1}{80} e^{-2 t} \left(120+136 \epsilon_1^2+3 \sqrt{2} \beta_1^2\right) +\cdots \non \\
&&-\frac{e^{-4 t} (6297+3584 \rho_4+432 t+\cdots )}{5376} +\cdots \nonumber\\
b_1(t) &=& e^{-t} \beta_1-\frac{1}{4} e^{-2 t} \left(3\ 2^{1/4} \sqrt{3}+4 \epsilon_1 \beta_1\right)+\cdots +\frac{1}{12} e^{-5 t}( m_3 +\cdots ) +\cdots \nonumber\\
b_2(t) &=& -2 e^{-t} \beta_1+\frac{1}{2} e^{-2 t} \left(3\ 2^{1/4} \sqrt{3}+10 \epsilon_1 \beta_1 \right)+\cdots +\frac{1}{12} e^{-5 t}( m_3 +\cdots ) +\cdots \nonumber\\
\rho(t) &=&
\frac{3}{40} e^{-2 t} \left(8 \epsilon_1^2-\sqrt{2} \beta_1^2\right) +\cdots + \frac{1}{224} e^{-4 t} (224 \rho_4+27 t +\cdots) +\cdots  \nonumber\\
\theta(t) &=& 
-\frac{15}{64} \sqrt{3} e^{-2 t}+\frac{9}{40} e^{-3 t} \left(3 \sqrt{3} \epsilon_1+2^{3/4} \beta_1\right)+\cdots   \nonumber\\ 
&+ &\!\!\!\!\! \frac{e^{-5 t} \! \left(12 \sqrt{3} \epsilon_1 (2529+3312 t)+2^{1/4} 7\left(160 \sqrt{2} m_3-9 \sqrt{2} \beta_1 (-157+264 t)\right) +\cdots \right)}{35840}+\cdots  \nonumber\\
q_0(t) &=&  
-\frac{15 \sqrt{3}}{8\ 2^{3/4}}+\frac{27}{5} e^{-t} \left(2^{1/4} \sqrt{3} \epsilon_1-\beta_1\right)+\cdots \non \\
&&+\frac{1}{140} e^{-3 t} \left(140 m_3+27 \left(92\ 2^{1/4} \sqrt{3} \epsilon_1-77 \beta_1\right) t\right)+\cdots \nonumber 
\end{eqnarray}
where the dots refer to exponentially suppressed terms in the expansion in $e^{-t}$ or to terms at least quadratic in the parameters $(\epsilon_1,\rho_4,\beta_1,m_3)$. 
As for the $Q^{111}$ black hole, we set two arbitrary constant terms  in the expansion of $u(t)$ and $v(t)$ to zero for notational simplicity; they can be restored applying the
transformations (\ref{simm1}) and (\ref{simm2}). 
The parameters $\epsilon_1$ and $\beta_1$ are associated with two modes with $\Delta=1$ belonging to the massless vector multiplet, while the parameters $\rho_4$ and $m_3$ correspond to a scalar with $\Delta=4$ and a  gauge mode with $\Delta=5$  in the massive vector multiplet (cfr  Table 7 of \cite{Cassani:2012pj}).  \\

Around the $AdS_2\times S^2$ vacuum there are four normalizable modes with behavior $e^{a \Delta t}$ with $\Delta = 0$, $\Delta=1$,  $\Delta=1.44$ and $\Delta=1.58$ where $a=4 \sqrt{2}$. At linear order the corresponding fluctuations are given by
modes $(U,V,v_1,v_2,b_1,b_2,\phi, \psi , q_0 )$ proportional to
\begin{eqnarray}
&& \{1,1,0,0,0,0,0,0,0\} 
 \nonumber \\
&&
\{-2.45,  -0.97, 
  1.22, 
  0.31, 
-0.09, 
  0.40, 
  0.82, \
-0.09, 
  1\} 
  \nonumber \\
 &&  \{0.05, 
  0.05, 
  0.30, 
-0.39, 
  -0.17, 
-0.64, 
  0.26, 
-0.41, 
  1\} 
  \nonumber\\
 && \{-0.27, 
-0.27, 
-1.85, 
  2.62, 
-4.81, 
-2.22, 
-1.23, 
-3.22, 
  1\}
 \end{eqnarray}
The mode with $\Delta=0$ is just a common shift in the metric functions corresponding to the symmetry (\ref{simm1}). The other modes give rise to a triple expansion in powers 
\be \sum_{p,q,r} d_{p,q,r} c_1^p c_2^q c_3 ^r  e^{(p + 1.44 q + 1.58 r) a  t}\ee of all the fields. 

\begin{figure}[h]
\begin{minipage}[t]{\linewidth}
~~~\begin{minipage}[t]{0.5\linewidth}
\vspace{0pt}
\centering
\includegraphics[scale=.6]{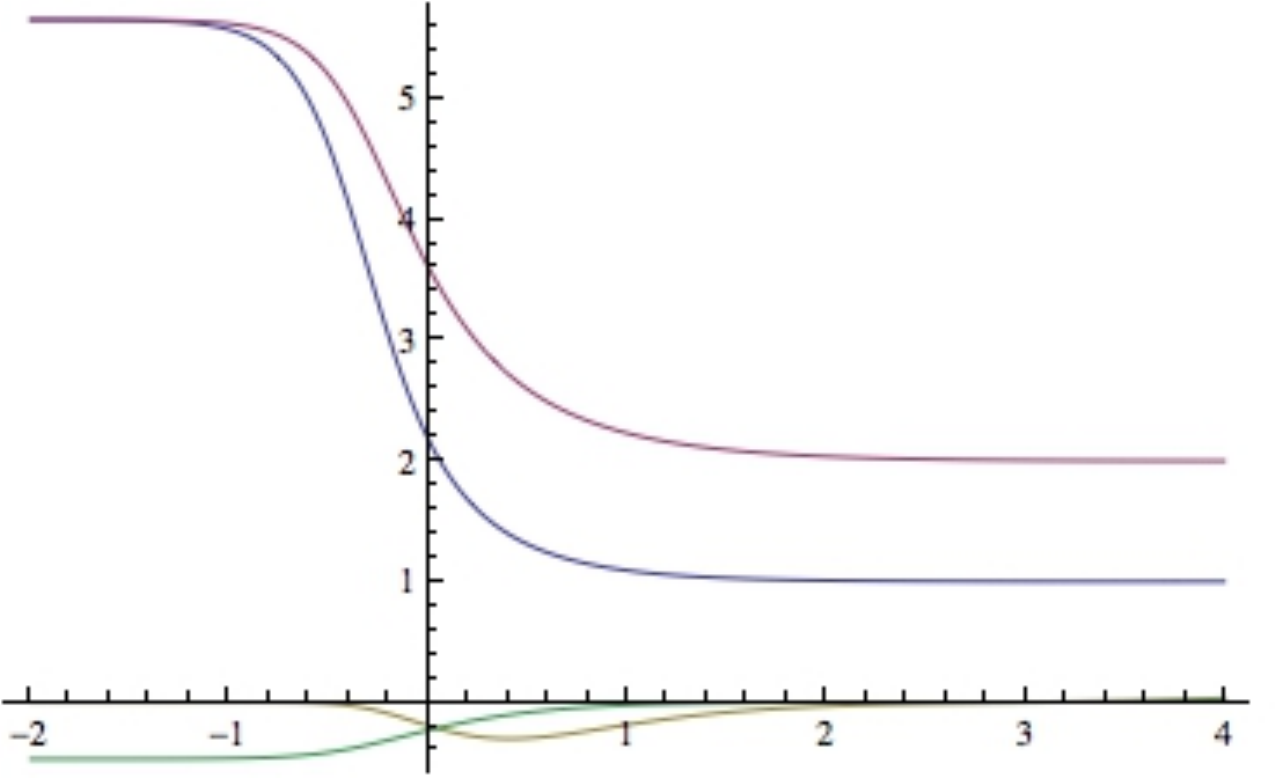}
\end{minipage}%
\hspace{.5cm}
\begin{minipage}[t]{0.5\linewidth}
\vspace{0pt}
\centering
\includegraphics[scale=.6]{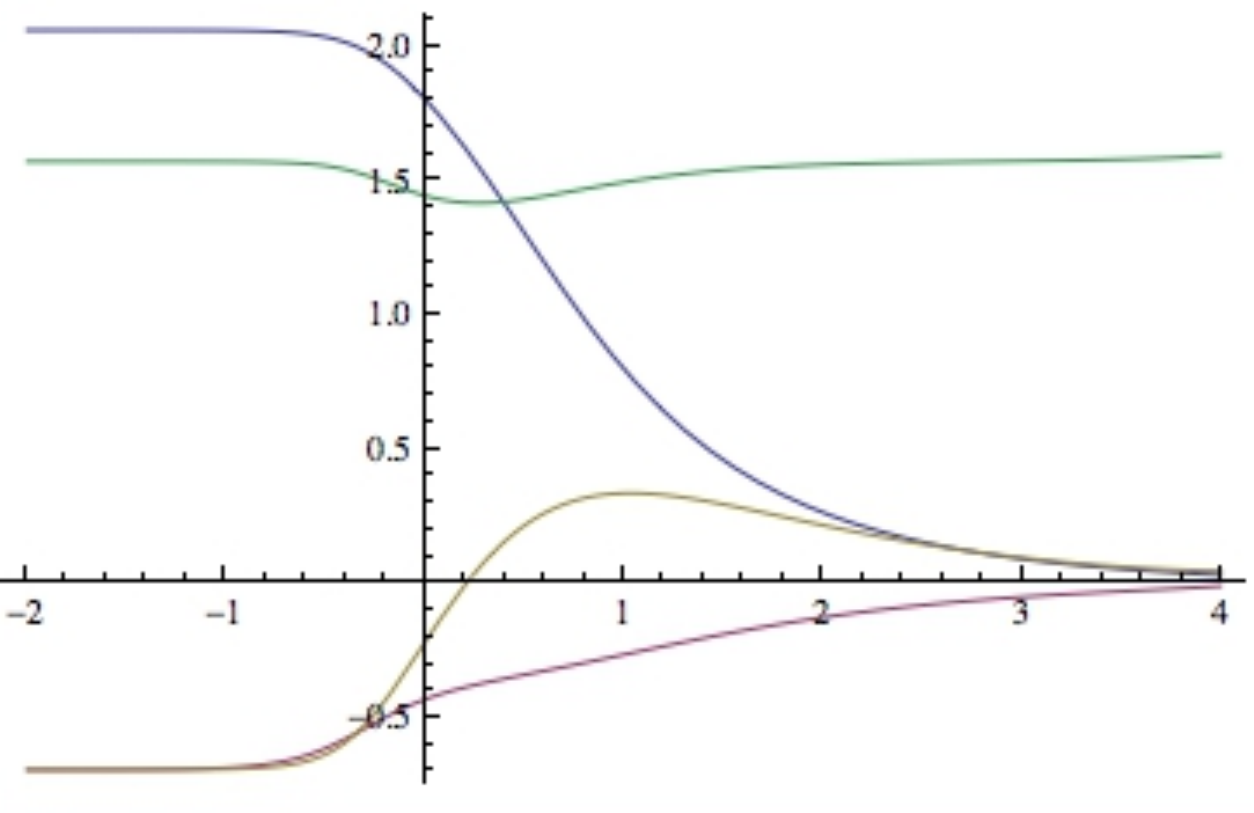}
\end{minipage}\\[1em]
\begin{minipage}[t]{0.5\linewidth}
\vspace{0pt}
\centering
\end{minipage}%
\vspace{1em}
\caption {Plots of $u',v', (2 b_1+b_2)/3,\rho$ on the left and of $(b_2-b_1)/3, e_1,e_2,\pi-\psi$ on the right corresponding to 
the value $c_1=1.7086,c_2=-2.4245,c_3=0.6713,c_4=-3.7021$. The UV expansion will be matched up to the
transformations (\ref{simm1}) and (\ref{simm2}).}
\label{fig:flowM110}
\end{minipage}
\end{figure}

 We have  a total number of eight parameters for nine equations which possess an algebraic integral of motion. We thus expect that the given IR and UV expansions can be matched at  finite $t$.  With some pain and using a precision much greater than the one given in the text above, we have 
 numerically solved the system of BPS equation and found an interpolating solution. The result is shown in Figure \ref{fig:flowM110}.
 
\vspace{1cm}

\noindent {\bf Acknowledgements}
A.Z.  is supported in part by INFN, the MIUR-FIRB grant RBFR10QS5J ``String Theory and Fundamental Interactions'', and by the MIUR-PRIN contract 2009-KHZKRX.
We would like to thank I. Bah, G. Dall'Agata, J. Gauntlett, K. Hristov, D. Klemm,  J. Simon and  B. Wecht for useful discussions and comments. 
M. P. would like to thank the members of the Theory Group at Imperial
College for their kind hospitality and support while this work was being completed.

\newpage

\begin{appendix}
\section{Four Dimensional Gauged Supergravity} \label{gsugra}

In this  Appendix, in order to fix notation and conventions, we recall few basic facts about  $\mathcal{N}=2$ gauged supergravity. 
We use the standard conventions of \cite{Andrianopoli:1996vr,Andrianopoli:1996cm}.

The fields of $\N=2$ supergravity are arranged into one graviton multiplet, $n_v$ vector multiplets
and $n_h$ hypermultiplets.  The graviton multiplet contains the metric, the graviphoton, $A_\mu^0 $  and an $SU(2)$ doublet of gravitinos of opposite chirality, ($ \psi_\mu^A, \psi_{\mu \, A} $),
where  $A=1,2$ is an $SU(2)$ index.  
The vector multiplets consist of a vector, $A^I_\mu,$, two spin 1/2 of opposite chirality, transforming as an $SU(2)$ doublet, ($\lambda^{i \,A}, \lambda^{\bar{i}}_A$), 
and one complex scalar $z^i$. $A=1,2$ is the $SU(2)$ index, while $I$ and $i$ run on the number
of vector multiplets $I= 1, \dots, n_{\rm V}$,  $i= 1, \dots, n_{\rm V}$.    Finally the hypermultiplets contain  two spin 1/2 fermions of opposite chirality,  ($\zeta_\alpha, \zeta^\alpha$),  and four real scalar fields, $q_u$,  where $\alpha = 1, \dots 2 n_{\rm H}$ and $u = 1, \ldots, 4 n_{\rm H}$.  
 
The scalars in the vector multiplets   parametrise a special K\"ahler manifold of  complex dimension $n_{\rm V}$, $\mathcal{M}_{\rm SK}$, 
with metric 
\be
g_{i \bar{j}} = - \del_i \del_{\bar{j}}  K(z, \bar{z}) 
\ee
where $ K(z, \bar{z})$ is the  K\"ahler potential on $\mathcal{M}_{\rm SK}$. This can be computed introducing 
homogeneous  coordinates $X^\Lambda(z)$ and define a holomorphic prepotential $\mathcal{F}(X)$, which is a homogeneous function of degree two
\be
\label{Kpotdef}
K(z \bar{z})  = - \ln  i (\bar{X}^\Lambda F_\Lambda - X^\Lambda \bar{F}_\Lambda)  \, ,
\ee
where $F_\Lambda = \del_\Lambda F$.  In the paper we will use both the holomorphic sections $(X^\Lambda, F_\Lambda)$ and  the symplectic sections 
\be
(L^\Lambda, M_\Lambda) =  e^{K/2}  (X^\Lambda, F_\Lambda) \, .
\ee

The scalars in the hypermultiplets   parametrise a quaternionic manifold of real dimension $4 n_{\rm H}$, $\mathcal{M}_{\rm Q}$, 
with metric $h_{uv}$.

The bosonic Lagrangian is 
\bea
\label{boslag}
\mathcal{L}_{\rm bos} & =&  -\frac{1}{2} R + i ( \bar{\cN}_{\Lambda \Sigma } \cF^{- \Lambda}_{\,\, \mu \nu} \cF^{- \Sigma \mu \nu}  -   \cN_{\Lambda \Sigma} \cF^{+ \Lambda}_{\,\, \mu \nu} \cF^{+ \Sigma \mu \nu} ) \non \\
&& + g_{i \bar{j}} \nabla^\mu z^i \nabla_\mu \bar{z}^{\bar{j}} + h_{u v} \nabla^\mu q^u \nabla_\mu q^{v}   - \mathcal{V}(z, \bar{z}, q)   \, ,
\eea
where $\Lambda, \Sigma = 0, 1, \ldots, n_{\rm V}$.  The gauge field strengths are defined as
\be
\cF^{\pm \Lambda}_{\mu\nu} = \half \blp F^{\Lambda}_{\mu\nu}\pm \frac{i}{2}\eps_{\mu\nu \rho \sig}F^{\Lambda \rho\sig} \brp,
\ee
with $F^\Lambda_{\mu \nu} = \frac{1}{2} (\partial_\mu A^\Lambda_\nu - \partial_\nu A^\Lambda_\mu)$.  In this notation, $A^0$ is
 the graviphoton  and $A^\Lambda$, with $\Lambda = 1, \ldots, n_{\rm V}$, denote  the vectors in the vector multiplets. 
The matrix $\cN_{\Lambda \Sigma}$ of the gauge kinetic term is a function of the vector multiplet scalars
\be
\label{periodmat}
\cN_{\Lambda \Sigma} = \bar{\mathcal{F}}_{\Lambda \Sigma} + 2 i \frac{\rm{Im} \mathcal{F}_{\Lambda \Delta} \rm{Im} \mathcal{F}_{\Lambda \Theta} X^\Delta X^\Theta}{ \rm{Im} \mathcal{F}_{ \Delta \Theta} X^\Delta X^\Theta}
\ee

The covariant derivatives  are defined as 
\bea
\label{scalarder}
&& \nabla_\mu z^i = \partial_\mu z^i + k^i_{\, \, \Lambda} A^{\Lambda}_{\, \, \mu} \, ,  \\
&& \nabla_\mu q^u = \partial_\mu q^u + k^u_{\, \, \Lambda} A^{\Lambda}_{\, \, \mu} \, ,
\eea
where $k^i_\Lambda$ and $k^u_\Lambda$ are the Killing vectors associated to the isometries of the vector and hypermultiplet scalar manifold that have been gauged. 
In this paper we will only gauge (electrically)  abelian  isometries of the hypermultiplet moduli space. 
The Killing vectors corresponding to quaternionic isometries have associated prepotentials: these are a set of real functions in the adoint of SU(2),
satisfying
\be
\Om^x_{uv}k^u_\Lam =-\nabla_v P^x_{\Lam} \, , 
\ee
where $\Om^x_{uv} = d \omega^x + 1/2 \epsilon^{x y z} \omega^y \wedge \omega^z$ and  $\nabla_v$ are the curvature and covariant derivative on $\cM_{{\rm Q}}$. In the specific models we consider in the text, one can show that the Killing vectors preserve the connection $\omega^x$ and the curvature $\Omega^x_{uv}$. This allows to simplify the prepotential
equations, which reduce to 
\be
P^x_\Lambda = k^u_\Lambda \omega^x_u \, .
\ee

Typically  in models obtained from $M$/string theory compactifications,  
the scalar fields have both electric and magnetic charges under the gauge symmetries.  However, by a symplectic transformation of the sections 
$(X^\Lambda, F_\Lambda)$, it is always  possible to put the theory in a frame where all scalars are electrically charged. Such a transformation\footnote{An $Sp( 2 + 2 n_{\rm V}, \mathbb{R})$ transformation of the sections 
 \be
 (X^\Lambda, F_\Lambda) \quad \mapsto \quad   (\tilde{X}^\Lambda, \tilde{F}_\Lambda) =    \bpm A& B \\C & D \epm (X^\Lambda, F_\Lambda) \, ,
 \ee
 acts on the period matrix
 $\cN_{\Lambda  \Sigma}$ by a fractional transformation
 \be
 \cN_{\Lambda \Sigma} (X, F)  \quad \mapsto  \quad   \tilde{\cN}_{\Lambda \Sigma} (\tilde{X}, \tilde{F}) = ( C + D   \cN_{\Lambda \Sigma} (X, F)) (A + B  \cN_{\Lambda  \Sigma} (X, F))^{-1} \, .
 \ee}
leaves the K\"ahler potential invariant, but changes the period matrix and the preprepotential $\mathcal{F}(X)$ . 

The models we consider in this paper \cite{Cassani:2012pj} are of this type: they have a cubic prepotential and  both electrical and magnetic gaugings
of some isometries of the hypermultiplet moduli space.  The idea is then to perform a sympletic rotation to a frame with purely electric gaungings,
allowing for sections $(\tX^{\Lam},\widetilde{F}_{\Lam})$ which are a general symplectic rotation of those
obtained from the cubic prepotential.

\vspace{0.2cm}

The scalar potential in \eqref{boslag} couples the hyper and vector multiplets,  and is given by
\be
\mathcal{V}(z, \bar{z}, q) = ( g_{i \bar{j}} k^i_\Lambda k^{\bar j}_\Sigma  
+ 4 h_{u v} k^u_\Lambda k^v_\Sigma ) \bar{L}^\Lambda L^\Sigma  + ( f_i^\Lambda g^{i \bar{j}} f^ \Sigma_{\bar j} - 3  \bar{L}^\Lambda L^\Sigma )
P^x_\Lambda P^x_\Sigma \, ,
\ee  
where $L^\Lambda$ are the symplectic sections  on $\mathcal{M}_{\rm SK}$, $f_i^\Lambda=   (\partial_i + \frac{1}{2}  \partial_i K)  L^\Lambda$ and $P^x_\Lambda$ are
the Killing prepotentials.

\vspace{0.4cm}

Maxwell's equation is
\be
\label{Maxwelleq}
\del_\mu \Blp \sqrt{-g} \blp \cI_{\Lam \Sig} F^{\Sig\, \mu \nu} + \half \cR_{\Lam \Sig}   \eps^{\mu\nu \rho \sig}F^{\Sig}_{\rho \sig}\brp  \Brp = \sqrt{-g}\,h_{uv} k^u_\Lam \nabla^\nu q^v
\ee
where, for simplicity of notation, we have defined  the following matrices
\be
\mathcal{R}_{\Lambda \Sigma} = {\rm Re} \mathcal{N}_{\Lambda \Sigma}  \qquad  \mathcal{I}_{\Lambda \Sigma} = {\rm Im} \mathcal{N}_{\Lambda \Sigma}  \, .
\ee

\vspace{0.4cm}

The full Lagrangian is invariant under $\cN =2$ supersymmetry. In the electric frame,  the  variations of  the fermionic fields  are given by
\bea
\label{gravitinoeq}
\delta \psi_{\mu A}&=& \mathcal{D}_\mu \epsilon_A + i   S_{AB} \gamma_\mu \epsilon^B  +  2i \, \cI_{\Lambda \Sigma} L^{\Sigma} \cF_{\mu\nu}^{- \Lambda} \gamma^\nu \eps_{AB}  \eps^B \, , \\
\label{gauginoeq}
\delta \lam^{iA}&=& i \nabla_\mu z^i \gam^\mu \eps^A -g^{i\jbar} \fbar^{\Sigma}_{\jbar}  \cI_{\Sig \Lam} \cF^{- \Lambda}_{\mu\nu} \gamma^{\mu\nu }\eps^{AB}\eps_B  + W^{i A B} \epsilon_B\, , \\
\label{hyperinoeq}
\delta \zeta_{\al}&=& i\, \cU^{B\beta}_u\nabla_\mu q^u \, \gamma^\mu \epsilon^A \epsilon_{AB}\epsilon_{\alpha\beta} + N^{A}_{\al} \eps_A  \, , 
\eea
where $ \cU^{B\beta}_u$ are the vielbeine on the quaternionic manifold and 
\bea
S_{AB}&=&\frac{i}{2} (\sigma_x)_A^{\phantom{A}C} \epsilon_{BC}
P^x_{\Lambda}L^\Lambda  \, , \nonumber\\
W^{iAB}&=&\epsilon^{AB}\,k_{\Lambda}^i \bar L^\Lambda\,+\,
{\rm i}(\sigma_x)_{C}^{\phantom{C}B} \epsilon^{CA} P^x_{\Lambda}
g^{ij^\star} {\bar f}_{j^\star}^{\Lambda}    \, ,    \label{pesamatrice}\\
{\cal N}^A_{\alpha}&=& 2 \,{\cal U}_{\alpha u}^A \,k^u_{\Lambda} \, 
\bar L^{\Lambda} \, .
\non
\eea

Notice that the  covariant derivative on the spinors 
\be
\mathcal{D}_\mu \epsilon_A =  \hD_{\mu}\eps_A  + \frac{i}{2}  (\sig^x)_A^{\ B} A^\Lam_{\mu} P^x_{\Lam} \eps_B \, .
\ee
contains a contribution from the gauge fields from the vector-$U(1)$ connection
\be
\hD_\mu \eps_A = (D_\mu+ \frac{i}{2} A_\mu)\eps_A +\omhat^x_{\mu}(\sig^x)_A^{\ B}\eps_B \, , \label{DepsA}
\ee
the hyper-$SU(2)$ connection and the gaugings (see eqs. 4.13,7.57, 8.5 in \cite{Andrianopoli:1996cm})
\bea
&& \omhat_\mu^x= \frac{i}{2} \del_\mu q^u \om^x_u \, , \\ 
&& A_\mu= \frac{1}{2i}(K_i \del_\mu z^i -K_{\ibar} \del_\mu z^{\ibar} ) \label{A1}\, .
\eea

\section{Derivation of the BPS Equations}
\label{sec:BPSEqs}

In this section we consider an ansatz for the metric and the gauge fields that  allows for black-holes with spherical or hyperbolic horizons, and we
derive the general conditions for 1/4 BPS solutions.  The metric and the gauge fields are taken to be 
\bea
\label{ansApp}
ds^2&=& e^{2U} dt^2- e^{-2U} dr^2- e^{2(V-U)} (d\tha^2+F(\tha)^2 d\vphi^2) \\
A^\Lam&=& \tq^\Lam(r) dt- p^\Lam(r) F'(\tha)  d\vphi \,, 
\eea
where the warp factors $U$ and $V$ are functions of the radial coordinate $r$ and
\be
F(\tha)=\left\{ \barr{ll} \sin \tha  & \qquad \quad  S^2\ (\kappa=1) \\
\sinh \tha  & \qquad \quad \HH^2\ (\kappa=-1) 
 \earr \right. 
\ee
The modifications needed for the flat case are discussed at the end of Section \ref{sec:BPSflow}.

We also assume that all scalars in the  vector and hypermultiplets, as well as the Killing spinors
$\epsilon_A$ are  functions of the radial coordinate only.

To derive the BPS conditions it is useful to introduce the central charge 
\bea
\cZ&=&p^\Lam M_\Lam- q_\Lam L^\Lam \non \\
&=& L^\Sig  \cI_{\Lam \Sig} (e^{2(V-U)} \tq^\Lam + i\kappa p^\Lam) \, ,
\eea
where $q_\Lam$ is defined in (\ref{maginv}) and its covariant derivative 
\be
D_{\ibar} \cZ =\fbar^{\Sigma}_{\ibar} \cI_{\Sig \Lam} \blp e^{2(V-U)} \tq'^{\Lam} +i\kappa p^{\Lam} \brp  \, .
\ee

In the case of flat space we need to replace $\kappa p^\Lam \rightarrow - p^\Lam$ in the definition (\ref{maginv}) of $q_\Lam$ and in the above expression for ${\cal Z}$.

\subsection{Gravitino Variation}

With the ansatz  \eqref{ansApp}, the gravitino variations  \eq{gravitinoeq} become 
\bea
 0&=&\frac{U'e^U}{2}\gam^{1}\eps_A 
+\frac{i}{2}  e^{-U}\tq^\Lam  P^x_\Lambda \, \gam^0(\sigma^x)_A^{\ B} \epsilon_B +iS_{AB}\eps^B  
 -\frac{i}{2} e^{2 (U- V)} \cM_+    \eps_{AB} \eps^B \label{gr1} \, , \\
0&=&\gam^1\hD_1\eps_A
+i S_{AB}\eps^B - \frac{i}{2} e^{2 (U- V)} \cM_-   \eps_{AB} \eps^B \label{gr2} \, , \\
0&=& \half (V'-U')e^U \gam^1\eps_A+
i S_{AB}\eps^B   + \frac{i}{2}  e^{2 (U- V)} \cM_-   \eps_{AB} \eps^B \label{gr3} \, , \\
0&=&\half e^{U-V} \frac{F'}{F} \gam^2 \eps_A+\half (V'-U')e^U \gam^1 \eps_A
-\frac{i}{2}  e^{U-V} \frac{F'}{F}p^\Lam  P^x_\Lambda \, \gam^3(\sigma^x)_A^{\ B} \epsilon_B 
+iS_{AB}\eps^B\non   \\&&
+\frac{i}{2} e^{2 (U- V)}   \cM_+   \eps_{AB} \eps^B 
  \label{gr4} \, ,
\eea
where, to simplify notations, we introduced the quantity
\be
\cM_\pm = \gamma^{01} \cZ \pm i \gamma^{02}  (F^{-1}   F^\prime \,  \cI_{\Lam \Sig} L^\Lam p'^\Sig) \, .
\ee

\vspace{0.2cm}

Let us consider first \eqref{gr1}.  The term proportional to $F^\prime$ must be separately zero, since it is the only 
$\tha$-dependent one. This implies
\be
\label{algc}
\cI_{\Lam \Sig} L^\Lam p'^\Sig  =0 \, .
\ee

Similarly, setting to zero the $\tha$-dependent terms in \eqref{gr4}, which is the usual statement of {\it setting the gauge connection equal to the spin connection},  gives the projector
\be
\label{proj1}
| \kappa|  \eps_A =   -  p^\Lam  P^x_\Lambda \,(\sigma^x)_A^{\ B} \gam^{01} \epsilon_B  \, .
\ee
This constraint also holds in the case of flat horizon if we set $\kappa=0$.
The $\tha$-independent parts of \eqref{gr4} and  \eqref{gr3} are equal and give
a second  projector
\be
\label{proj2}
S_{AB}\eps^B  = 
\frac{ i}{2}  (V'-U')e^U \gam^1\eps_A - \frac{1}{2}  e^{2 (U-V)} \cZ\gam^{01}   \eps_{AB} \eps^B   \, .
\ee
Subtracting the $\tha$ independent parts of \eqref{gr1} and \eqref{gr3} gives  a third projector
\be
\label{proj3}
\eps_A= - \frac{2i}{(2U'-V') } \Bslb  e^{U-2V} \cZ\eps_{AB}\gam^{0} \eps^B + \frac{1}{2} e^{-2U}  \tq^\Lam  P^x_\Lambda \, \gam^{01}(\sigma^x)_A^{\ B} \epsilon_B \Bsrb \, .
\ee

Finally, subtracting \eqref{gr2} and \eqref{gr1} we obtain an equation for the radial dependence of the spinor
\be
\hD_1 \eps_A=\frac{U'e^U}{2}\eps_A 
+ \frac{i}{2}  e^{-U}\tq^\Lam  P^x_\Lambda \, \gam^{01}(\sigma^x)_A^{\ B} \epsilon_B\,. \label{raddep}
\ee

In total we get three projectors, \eqref{proj1} - \eqref{proj3}, one differential relation on the spinor  \eqref{raddep} and one algebraic constraint \eqref{algc}. 
The idea is to further simply these equations so as to ensure that we end up with  two projectors.  From now on we will specify to 
the case of spherical or hyperbolic symmetry, since this is what we will use in the paper. In order to reduce the number of projectors we impose the  constraint
\be
\tq^\Lam P_\Lam^x = c\, e^{2U} \,  p^\Lam P_\Lam^x\,, \ \ x=1,2,3   \label{pcq}
\ee
for some real function $c$. By squaring \eqref{proj1} we obtain the algebraic condition
\be
(p^\Lam P_\Lam^x)^2=  1 
 \ee
which can be used to rewrite \eqref{pcq} as
\be
\label{ceq}
c=   e^{-2U}\tq^\Lam P_\Lam^x p^\Sig P_\Sig^x\,.
\ee

Substituting \eqref{proj1} in \eqref{proj3} and using \eqref{ceq}, we obtain the projector
\be
\label{projNa}
\epsilon_A= -  \frac{2 i e^{U-2 V} \cZ }{2 U^\prime - V^\prime - i c} \epsilon_{A B} \gamma^0 \epsilon^B
\ee
which, squared, gives the norm of $\cZ$
\bea
|\cZ|^2=\frac{1}{4} e^{4V - 2U}[(2U'-V')^2 +c^2] \, . \label{normZ}
\eea

Then we can rewrite \eqref{projNa} as 
\be
 \eps_A=ie^{i \psi}  \eps_{AB}\gam^{0}  \eps^B  
 \label{proj4} \, ,
 \ee
where $e^{i \psi}$ is the relative phase between $\cZ$ and $2 U^\prime - V^\prime - i c$
\be
\label{phase}
e^{i \psi} = -  \frac{2 e^{U-2V} \cZ }{2 U^\prime - V^\prime - i c } \, .
\ee
Using the definition of $S_{AB}$ given in \eqref{pesamatrice} and the projectors
\eqref{proj1} and \eqref{proj4}, we can reduce \eqref{pesamatrice} to a scalar equation
\be
i \cL^\Lam P^x_{\Lambda} p^\Sig P_\Sig^x = \Bslb e^{2(U-V)} \cZ  e^{- i\psi} -(V'-U')e^{U} \Bsrb    \,, 
 \label{LPrel}
\ee
where we defined
\be
\cL^{\Lam}= e^{- i\psi}L^\Lam=\cL_r+ i \cL_i\,.
\ee
Combining \eqref{phase} and \eqref{normZ}, we can also write two equations for the warp factors
\bea
&& e^U U'=  -i \cL^\Lam P^x_{\Lambda} p^\Sig P_\Sig^x   - e^{2 (U- V)}  \cZ  e^{-i \psi}   + i c e^{U} \, ,   \\
&& e^U V'  =  -2i  \cL^\Lam P^x_{\Lambda} p^\Sig P_\Sig^x + i c e^{U} \, . 
\eea

Using the projectors above, \eqref{raddep}  becomes
\be
\del_r \eps_A=-\frac{i}{2}A_r\eps_A -\omhat^x_{r}(\sig^x)_A^{\ B}\eps_B+\frac{U'}{2}\eps_A 
- \frac{ic}{2}\,\epsilon_A\,. 
\ee

\subsection{Gaugino Variation}

The gaugino variation is
\be
\label{zdot1}
 i e^{U} z'^i \gam^1 \eps^A  +e^{2 (U-V)} g^{i \bar{j}}  \big[ D_{\ibar} \cZ  \gamma^{01} - (F^{-1}   F^\prime \, \fbar^{\Sigma}_{\jbar} \cI_{\Sig \Lam} p'^{\Lam} ) \gamma^{13} \big]
  \eps^{AB}\epsilon_B  + W^{i A B} \epsilon_B  = 0  \,  .
\ee
$\cM$ is the only  $\tha$-dependent term and  must be set to zero separately, giving
\be
\label{Bpfbar}
\fbar^{\Sigma}_{\jbar} \cI_{\Sig \Lam} p'^{\Lam}= 0  \,.
\ee

Combining  \eq{algc} and  \eq{Bpfbar}, and using  standard orthogonality relations between the sections $X^\Lambda$, we conclude that 
\be
p'^{\Lam}=0 \, .
\ee

Continuing with \eq{zdot1},  we use again \eqref{proj1} and  \eq{proj4} to obtain
\be
\label{gauge3}
 e^{-i \psi} e^{U} z'^i   =  e^{2 (U-V)} g^{i \bar{j}}   D_{\ibar} \cZ 
 - i  g^{i\jbar} {\bar f}_{\jbar}^{\Lambda} P_\Lam^xp^\Sig P_\Sig^x \, .
\ee
%

\subsection{Hyperino Variation}
The hyperino variation gives
\bea
i\,   \eps_{\al \beta} \cU^{B\beta}_u
\blp e^U \gam^1  q'^u +  \tq^\Lam  k^u_{\Lam } e^{-U} \gam^0 -F^{-1}F' e^{U-V} p^\Lam  k^u_{\Lam }\gam^3\brp 
\eps_{AB}\eps^A+ 2 \,{\cal U}_{\alpha u}^A \,k^u_{\Lam} 
\Lbar^{\Lam} \eps_A = 0 \, .
\eea
First off, we need to set the $\tha$-dependent part to zero
\be
 k_{\Lam}^u p^\Lam = 0 \,. \label{kqzero1} 
\ee
The projectors \eqref{proj1} and \eqref{proj4} can be used to simply  the remaining equation 
\be
-  e^{U}  q'^u  \, \cU^{B}_{\alpha \, u}  p^\Lambda P^x_\Lambda (\sigma^x)_B^C \epsilon_C + 
\cU^{A}_{\alpha \, u} ( 2   k^u_{\Lam} \Lbar^{\Lam} - e^{-U}  \tq^\Lam  k^u_{\Lam })  \eps^A = 0 \, ,
\ee
which can then be reduced to a scalar equation
\be
\label{hyper1}
-i   h_{uv} q'^u 
+e^{-2U}p^\Sig P^y_\Sig \tq^\Lam\nabla_v  P^y_\Lam
- 2e^{-U}  p^\Sig P^x_\Sig \nabla_v (\ol{\cL}^{\Lam} P^x_\Lam) = 0 \, .
\ee
Using the standard relations (we use the conventions of \cite{Andrianopoli:1996vr}) 
\bea
 -i\Om_{u}^{x\,v} \cU_{v}^{A\al}&=&   \cU^{B\al}_{u}(\sig^x)_B^{\ A}  \, , \non \\
 \Om^x_{uw} \Om^{y\,w}_{\ \ v}&=& -\delta^{xy}h_{uv}-\eps^{xyz}\Om^z_{uv} \, , \\
 k^u_\Lam \Om^x_{uv}&=&- \nabla_v P^x_\Lam\, , 
\eea
we can reduce \eqref{hyper1} to
\be
-i   h_{uv} q'^u 
+e^{-2U}p^\Sig P^y_\Sig \tq^\Lam\nabla_v  P^y_\Lam
- 2e^{-U}  p^\Sig P^x_\Sig \nabla_v (\ol{\cL}^{\Lam} P^x_\Lam) =0  \, .
\ee
The real and imaginary parts give
\be
\barr{rcl}
 q'^u &=&2e^{-U}   h^{uv}\del_v \blp p^\Sig P^x_\Sig \cL_i^{\Lam} P^x_\Lam\brp \, ,  \\
0&=&  \tq^\Lam  k^u_{\Lam }  - 2 e^U \cL_r^{\Lam}k^u_{\Lam}  \, .
\earr 
\ee

\subsection{Summary of BPS Flow Equations}

It is worthwhile at this point to summarize the BPS equations. The algebraic equations are
\bea
p'^\Lam&=& 0 \label{BPS1} \, , \\
( p^\Lam P_\Lam^x)^2&=&1 \, ,   \label{BPS2}\\
k_{\Lam}^u p^\Lam &=& 0 \, , \label{BPS3} \\
\tq^\Lam P^x_\Lam&=& c \,e^{2U} p^\Lam P^x_\Lam \, , \label{BPS4}\\
\tq^\Lam  k^u_{\Lam }  &=& 2 e^U \cL_r^{\Lam}k^u_{\Lam} \, ,  \label{BPS5}
\eea
while the differential equations are
\bea
 e^U U'&=& -i \cL^\Lam P^x_{\Lambda} p^\Sig P_\Sig^x   +  \cN e^{-i \psi}   + i c e^{U} \, ,   \\
 e^U V'  &=&  -2i  \cL^\Lam P^x_{\Lambda} p^\Sig P_\Sig^x + i c e^{U} \, ,  \\
 e^{-i \psi} e^{U} z'^i   &=&  \cN^i 
 - i  g^{i\jbar} {\bar f}_{\jbar}^{\Lambda} P_\Lam^xp^\Sig P_\Sig^x  \, , \label{gaugino}\\
  q'^u&=&2e^{-U}  h^{uv}\del_v \Blp p^\Sig P^x_\Sig \cL_i^{\Lam} P^x_\Lam\Brp  \, . 
\eea

In the case of flat horizon  equation (\ref{BPS2}) is replaced by $( p^\Lam P_\Lam^x)^2=0$.

\subsection{Maxwell's Equation}

Maxwell's equation is
\be
\del_\mu \Blp \sqrt{-g} \blp \cI_{\Lam \Sig} F^{\Sig\, \mu \nu} + \half \cR_{\Lam \Sig}   \eps^{\mu\nu \rho \sig}F^{\Sig}_{\rho \sig}\brp  \Brp = \sqrt{-g}\,h_{uv} k^u_\Lam \nabla^\nu q^v \, , 
\ee
which gives
\be
q'_\Lam\equiv \Blp -e^{2(V-U)} \cI_{\Lam\Sig} \tq'^\Sig +\cR_{\Lam\Sig} \kappa  p^\Sig \Brp'=  2 e^{2V-4U}h_{uv} k^u_\Lam k^v_\Sig \tq^\Sig
\ee

In the case of flat horizon we need to replace $\kappa p^\Lam\rightarrow - p^\Lam$.

\end{appendix}


\providecommand{\href}[2]{#2}\begingroup\raggedright\endgroup


\end{document}